\input ./style/arxiv-general.cfg
\documentclass[aap,MSNbibl,seceqn,dvips]{arximspdf}
\makeatletter
   \@ifpackageloaded{graphicx}{}{\usepackage{graphicx}}
\makeatother
\doi{10.1214/14-AAP1059} % kopijuoti is laisko
\volume{25}
\issue{5}
\pubyear{2015}
\firstpage{2708}
\lastpage{2742}
\docsubty{FLA}

\makeatletter
\newcommand{\ucp}{\mathrm{ucp}}
\newcommand{\rrvert}{\vert}
\newcommand{\rrVert}{\Vert}
\newcommand{\llvert}{\vert}
\newcommand{\llVert}{\Vert}
\renewcommand{\mid}{|}
\newtheorem{Theorem}{Theorem}[section] %
\newtheorem{Corollary}[Theorem]{Corollary}%
\newtheorem{Lemma}[Theorem]{Lemma} %
\newproclaim{Assumption}[Theorem]{Assumption}%
\newproclaim{Definition}[Theorem]{Definition}%
\newproclaim{Example}[Theorem]{Example} %
\newproclaim{Question}[Theorem]{Question}%
\newproclaim{Remark}[Theorem]{Remark} %
\newcommand{\idvec}{\mathbf{1}}
\newcommand{\set}{\triangleq}
\newcommand{\setind}[1]{1_{#1}}
\newcommand{\const}{\operatorname{const}}
\makeatother

\begin{document}
\begin{frontmatter}

\title{A model for a large investor trading at market indifference
prices. II: Continuous-time case}
\runtitle{Continuous-time model for a large investor}

\begin{aug}
\author[A]{\fnms{Peter} \snm{Bank}\thanksref{T1}\ead[label=e1]{bank@math.tu-berlin.de}}
\and
\author[B]{\fnms{Dmitry}~\snm{Kramkov}\corref{}\thanksref{T2}\ead[label=e2]{kramkov@cmu.edu}}
\runauthor{P. Bank and D. Kramkov}
\affiliation{Technische Universit{\"a}t Berlin and Carnegie Mellon University}
%\dedicated{}
\address[A]{Institut f{\"u}r Mathematik\\
Technische Universit{\"a}t Berlin\\
Stra{\ss}e des 17. Juni 136\\
10623 Berlin\\
Germany\\
\printead{e1}}
\address[B]{Department of Mathematical Sciences\\
Carnegie Mellon University\\
5000 Forbes Avenue\\
Pittsburgh, Pennsylvania 15213-3890\\
USA\\
\printead{e2}}
\end{aug}
\thankstext{T1}{Supported in part by NSF
under Grant DMS-05-05021.}
\thankstext{T2}{The author also holds a part-time position at the
University of Oxford. Supported in part by
NSF under Grant DMS-05-05414 and by the Carnegie Mellon Portugal Program.}

% HISTORY:
\received{\smonth{9} \syear{2011}}
\revised{\smonth{1} \syear{2014}}
%\accepted{\smonth{} \syear{}}

% ABSTRACT
%
\begin{abstract}
We develop from basic economic principles a continuous-time model
for a large investor who trades with a finite number of market
makers at their utility indifference prices. In this model, the
market makers compete with their quotes for the investor's orders
and trade among themselves to attain Pareto optimal allocations. We
first consider the case of simple strategies and then, in analogy to
the construction of stochastic integrals, investigate the transition
to general continuous dynamics. As a result, we show that the
model's evolution can be described by a nonlinear stochastic
differential equation for the market makers' expected utilities.
\end{abstract}

% KEYWORDS
\end{frontmatter}

%s1 #&#
\section{Introduction}\label{secintroduction}

A typical financial model presumes that the prices of traded
securities are not affected by an investor's buy and sell orders.
From a practical viewpoint, this assumption is justified as long as his
trading volume remains small enough to be easily covered by market
liquidity. An opposite situation occurs, for instance, when an
economic agent has to sell a large block of shares over a short period
of time; see, for example, Almgren and Chriss \cite{AlmgChr01} and Schied and Sch{\"o}neborn \cite{SchSchon09}. This
and other examples motivate the development of financial models for a
``large'' trader, where the dependence of market prices on his
strategy, called a \emph{price impact} or a \emph{demand pressure}, is
taken into account.

Hereafter, we assume that the interest rate is zero and, in
particular, is not affected by the large investor. As usual in
mathematical finance, we describe a \mbox{(self-financing)} strategy by a
predictable process $Q=(Q_t)_{0\leq t\leq T}$ where $Q_t$ is the
number of stocks held \emph{just before} time $t$ and $T$ is a finite
time horizon. The role of a ``model'' is to define a predictable
process $X(Q)$ representing the evolution of the cash balance for the
strategy $Q$. We denote by $S(Q)$ the \emph{marginal} price process
of traded stocks, that is, $S_t(Q)$ is the price at which one can
trade an infinitesimal quantity of stocks at time~$t$. Recall that in
the standard model of a ``small'' agent the price $S$ does not depend
on $Q$ and
\[
X_t(Q) = \int_0^t Q_u
\,dS_u - Q_t S_{t}.
\]

In mathematical finance, a common approach is to specify the price
impact of trades \emph{exogenously}, that is, to postulate it as one
of the inputs. For example, Frey and Stremme \cite{FreyStr97},
Platen and Schweizer \cite{PlatSch98},
Papanicolaou and Sircar \cite{PapanSirc98} and
Bank and Baum \cite{BankBaum04}
choose a stochastic field of \emph{reaction functions}, which
explicitly state the dependence of the marginal prices on the
investor's current holdings, {\c{C}}etin, Jarrow and Protter  in
\cite{CetinJarrProt04} start with a stochastic field of \emph{supply
curves}, which define the prices in terms of traded quantities
(\emph{changes} in holdings), and Cvitani{\'c} and Ma \cite{CvitMa96}
make the drift and the volatility of the price
process dependent on a trading strategy; we refer the reader to the
recent survey \cite{GokayRochSoner11} by
G{\"o}kay, Roch and Soner for more details and additional
references. Note that in all these models the processes $X(Q)$ and
$S(Q)$, of the cash balance and of the marginal stock price, only
depend on the ``past'' of the strategy $Q$, in the sense that
%
%e1.1 #&#
\begin{equation}
\label{eq1} X_t(Q) = X_t\bigl(Q^t\bigr),
\qquad S_t(Q) = S_t\bigl(Q^t\bigr),
\end{equation}
where $Q^t \set(Q_{s\wedge t})_{0\leq s\leq T}$ denotes the process
$Q$ ``stopped'' at $t$ with $s\wedge t \set\min(s,t)$.

The exogenous nature of the above models facilitates their calibration
to market data; see, for example, \cite{CetinJarrProtWarac06} by
{\c{C}}etin, Jarrow and Protter. There are, however, some
disadvantages. For example, the models in \cite{FreyStr97,PlatSch98,PapanSirc98,BankBaum04,CetinJarrProt04} and \cite{CetinJarrProtWarac06} do not
satisfy the natural ``closability'' property for a large investor
model:
%
%e1.2 #&#
\begin{equation}
\label{eq2} \bigl\llvert Q^n\bigr\rrvert \leq\frac{1}n
\quad\Longrightarrow\quad X_T\bigl(Q^n\bigr) \to 0, \qquad n\to
\infty,
\end{equation}
while in Cvitani{\'c} and Ma \cite{CvitMa96} the stock price is not affected by a jump in
investor's holdings: $S_t(Q_t + \Delta Q_t) = S_t(Q_t)$.

In our project, we seek to derive the dependence of prices on
strategies \emph{endogenously} by relying on the framework developed
in financial economics. A starting point here is the postulate that,
at any given moment, a price reflects a balance between demand and
supply or, more formally, it is an output of an \emph{equilibrium}. In
addition to the references cited below, we refer the reader to the book
\cite{Hara95} by O'Hara and the survey
\cite{AmihudMendPeder05} by Amihud, Mendelson and Pedersen.

To be more specific, denote by $\psi$ the terminal price of the traded
security, which we assume to be given exogenously, that is, $S_T(Q) =
\psi$ for every strategy $Q$. Recall that in a small agent model the
absence of arbitrage implies the existence of an equivalent
probability measure $\mathbb{Q}$ such that
%
%e1.3 #&#
\begin{equation}
\label{eq3} S_t = \mathbb{E}_{\mathbb{Q}}[\psi\mid
\mathcal{F}_t], \qquad0\leq t\leq T,
\end{equation}
where $\mathcal{F}_t$ is the $\sigma$-field describing the information
available at time $t$. This result is often called \emph{the
fundamental theorem of asset pricing}; in full generality, it has
been proved by Delbaen and Schachermayer  in \cite{DelbSch94,DelbSch98}. The economic nature of this \emph{pricing measure}
$\mathbb{Q}$ does not matter in the standard, small agent,
setup. However, it becomes important in an equilibrium-based
construction of models for a large trader where it typically
originates from a Pareto optimal allocation of wealth and is given by
the expression~(\ref{eq4}) below.

We shall consider an economy formed by $M$ market participants, called
hereafter the \emph{market makers}, whose preferences for terminal
wealth are defined by utility functions $u_m=u_m(x)$, $m=1,\ldots,M$,
and an identical subjective probability measure $\mathbb{P}$. It is
well known in financial economics that the Pareto optimality of the
market makers' wealth allocation $\alpha=(\alpha^m)_{m=1,\ldots,M}$
yields the pricing measure~$\mathbb{Q}$ defined by
%
%e1.4 #&#
\begin{equation}
\label{eq4} \frac{d\mathbb{Q}}{d\mathbb{P}} = v^m u_m'
\bigl(\alpha^m\bigr), \qquad m=1,\ldots,M,
\end{equation}
where $v^m>0$ is a normalizing constant.

It is natural to expect that in the case when the strategy $Q$ is not
anymore negligible an expression similar to~(\ref{eq3}) should still
hold true for the \emph{marginal} price process:
%
%e1.5 #&#
\begin{equation}
\label{eq5} S_t(Q) = \mathbb{E}_{\mathbb{Q}_t(Q)}\bigl[\psi\mid
\mathcal{F}_t(Q)\bigr], \qquad 0\leq t\leq T.
\end{equation}
This indicates that the price impact at time $t$ described by the
mapping $Q \mapsto S_t(Q)$ may be attributed to two common aspects of
market's microstructure:
\begin{longlist}
\item[1. \emph{Information}:] $Q\mapsto\mathcal{F}_t(Q)$. Models focusing
on information aspects naturally occur in the presence of an
insider, where $\mathcal{F}_t(Q)$, the information available to the
market makers at time $t$, is usually generated by the sum of $Q$
and the cumulative demand process of ``noise'' traders; see
Glosten and Milgrom \cite{GlosMilg85},
Kyle \cite{Kyle85} and
Back and Baruch \cite{BackBaruch04},
among others.
\item[2. \emph{Inventory}:] $Q \mapsto\mathbb{Q}_t(Q)$. In view
of~(\ref{eq4}), this reflects how $\alpha_t(Q)$, the Pareto optimal
allocation of the total wealth or ``inventory'' induced by $Q$,
affects the valuation of marginal trades. Note that the random
variable $\alpha_t(Q)$ is measurable with respect to the terminal
$\sigma$-field $\mathcal{F}_T(Q)$ [not with respect to the current
$\sigma$-field $\mathcal{F}_t(Q)$!].
\end{longlist}

In our study, we shall focus on the inventory aspect of price formation
and disregard the informational component. We assume that the market
makers share the same exogenously given filtration
$(\mathcal{F}_t)_{0\leq t\leq T}$ as the large trader and, in
particular, their information flow is not affected by his strategy
$Q$:
\[
\mathcal{F}_t(Q) = \mathcal{F}_t, \qquad0\leq t\leq T.
\]
Note that this informational symmetry is postulated only regarding the
externally given random outcome. As we shall discuss below, in
inventory based models, the actual form of the map $Q \mapsto
\mathbb{Q}_t(Q)$, or, equivalently, $Q\mapsto\alpha_t(Q)$ is implied
by game-theoretical features of the interaction between the market
makers and the investor. In particular, it depends on the knowledge
the market makers possess at time $t$ about the subsequent evolution
$(Q_s)_{t\leq s\leq T}$ of the investor's strategy, conditionally on
the forthcoming random outcome on $[t,T]$.

For example, the models in
Grossman and Miller \cite{GrosMill88},
Garleanu, Pedersen and Poteshman \cite{GarlPederPotes09} and
German \cite{Germ11} rely on a setup
inspired by the \emph{Arrow--Debreu equilibrium}. Their framework
implicitly assumes that right from the start the market makers have
\emph{full knowledge} of the investor's future strategy $Q$ (of
course, contingent on the unfolding random scenario). In this case,
the resulting pricing measures and the Pareto allocations do not
depend on time:
%
%e1.6 #&#
\begin{equation}
\label{eq6} \mathbb{Q}_t(Q) = \mathbb{Q}(Q), \qquad
\alpha_t(Q) = \alpha(Q), \qquad0\leq t\leq T,
\end{equation}
and are determined by the budget equations:
\[
\mathbb{E}_{\mathbb{Q}(Q)}\bigl[\alpha^m(0)\bigr] =
\mathbb{E}_{\mathbb{Q}(Q) }\bigl[\alpha^m(Q)\bigr], \qquad m=1,\ldots,M,
\]
and the clearing condition:
\[
\sum_{m=1}^M \alpha^m(Q) =
\sum_{m=1}^M \alpha^m(0) + \int
_0^T Q_t\,d {S_t(Q)}.
\]
Here, $\mathbb{Q}(Q)$ and $S(Q)$ are defined in terms of $\alpha(Q)$
by~(\ref{eq4}) and~(\ref{eq5}). The positive sign in the clearing
condition is due to our convention to interpret $Q$ as the number of
stocks held by the market makers. It is instructive to note that in
the case of exponential utilities, when $u_m(x)=-\exp(-a_m x)$ with a
risk-aversion $a_m>0$, the stock price in these models depends only on
the ``future'' of the strategy:
\[
S_t(Q) = S_t\bigl((Q_s)_{t\leq s\leq T}
\bigr), \qquad0\leq t\leq T,
\]
which is just the opposite of~(\ref{eq1}).

In our model, the interaction between the market makers and the
investor takes place according to a \emph{Bertrand competition}; a
similar framework (but with a single market maker and only in a
one-period setting) was used in Stoll \cite{Stoll78}. The key economic
assumptions can be summarized as follows:
\begin{longlist}[2.]
\item[1.] After every trade, the market makers can redistribute new income
to form a Pareto allocation.
\item[2.] As a result of a trade, the expected utilities of the market
makers do not change.
\end{longlist}
The first condition assumes that the market makers are able to find
the most effective way to share among themselves the risk of the
resulting total endowment, thus producing a Pareto optimal
allocation. The second assumption is a consequence of a Bertrand
competition which forces the market makers to quote the most
aggressive prices without lowering their expected utilities; in the
limit, these utilities are left unchanged.

Our framework implicitly assumes that at every time $t$ the market
makers have \emph{no a priori knowledge} about the subsequent trading
strategy $(Q_s)_{t\leq s\leq T}$ of the economic agent (even
conditionally on the future random outcome). As a consequence, the
marginal price process $S(Q)$ and the cash balance process $X(Q)$ are
related to $Q$ as in~(\ref{eq1}). Similarly, the dependence on $Q$ of
the pricing measures and of the Pareto optimal allocations is
nonanticipative in the sense that
\[
\mathbb{Q}_t(Q) = \mathbb{Q}_t\bigl(Q^t
\bigr), \qquad\alpha_t(Q) = \alpha_t\bigl(Q^t
\bigr), \qquad0\leq t\leq T,
\]
which is quite opposite to~(\ref{eq6}).

In \cite{BankKram13a}, we studied the model in a static, one-step,
setting. The current paper deals with the general continuous-time
framework. Building on the single-period case in an inductive manner,
we first define \emph{simple strategies}, where the trades occur only
at a finite number of times; see Theorem~\ref{th1}. The main
challenge is then to show that this construction allows for \emph{a
consistent passage} to general predictable strategies. For instance,
it is an issue to verify that the cash balance process $X(Q)$ is
stable with respect to uniform perturbations of the strategy $Q$ and,
in particular, that the closability property~(\ref{eq2}) and its
generalizations stated in Questions~\ref{quest1} and \ref{quest2}
hold.

These stability questions are addressed by deriving and analyzing a
nonlinear stochastic differential equation for the market makers'
expected utilities; see~(\ref{eq59}) in Theorem~\ref{th4}. A key
role is played by the fact, that together with the strategy $Q$, these
utilities form a ``sufficient statistics'' in the model, that is, they
uniquely determine the Pareto optimal allocation of wealth among the
market makers. The corresponding functional dependencies are
explicitly given as gradients of the \emph{stochastic field of
aggregate utilities} and its saddle conjugate; here we rely on our
companion paper~\cite{BankKram13b}.

An outline of this paper is as follows. In Section~\ref{secmodel}, we
define the model and study the case when the investor trades according
to a simple strategy. In Section~\ref{secstoch-processes}, we provide
a conditional version of the well-known parameterization of Pareto
optimal allocations and recall basic results from \cite{BankKram13b}
concerning the stochastic field of aggregate utilities and its
conjugate. With these tools at hand, we formally define the
strategies with general continuous dynamics in
Section~\ref{secstrat-with-cont}. We conclude with
Section~\ref{secappr-simple-strat} by showing that the construction
of strategies in Section~\ref{secstrat-with-cont} is consistent with
the original idea based on the approximation by simple strategies. In
the last two sections, we restrict ourselves to a Brownian setting, due
to convenience of references to Kunita~\cite{Kunit90}.

%s2 #&#
\section{Model}\label{secmodel}

%s2.1 #&#
\subsection{Market makers and the large investor}\label{secfinancial-market}

We consider a financial model where $M\in \lbrace 1,2,\ldots \rbrace$ market
makers quote prices for a finite number of stocks. Uncertainty and
the flow of information are modeled by a filtered probability space
$(\Omega, \mathcal{F}, (\mathcal{F}_t)_{0 \leq t \leq T}, \mathbb{P})$
satisfying the standard conditions of right-continuity and
completeness; the initial $\sigma$-field $\mathcal{F}_0$ is trivial,
$T$ is a finite maturity and $\mathcal{F} = \mathcal{F}_T$.

As usual, we identify random variables differing on a set of
$\mathbb{P}$-measure zero; $\mathbf{L}^0(\mathbf{R}^d)$ stands for the
metric space of such equivalence classes with values in $\mathbf{R}^d$
endowed with the topology of convergence in probability;
$\mathbf{L}^p(\mathbf{R}^d)$, $p\geq1$, denotes the Banach space of
$p$-integrable random variables. For a $\sigma$-field
$\mathcal{A}\subset\mathcal{F}$ and a set $A\subset\mathbf{R}^d$
denote $\mathbf{L}^0(\mathcal{A},A)$ and
$\mathbf{L}^p(\mathcal{A},A)$, $p\geq1$, the respective subsets of
$\mathbf{L}^0(\mathbf{R}^d)$ and $\mathbf{L}^p(\mathbf{R}^d)$
consisting of all $\mathcal{A}$-measurable random variables with
values in $A$.

The way the market makers serve the incoming orders crucially depends
on their attitude toward risk, which we model in the classical
framework of expected utility. Thus, we interpret the probability
measure $\mathbb{P}$ as a description of the common beliefs of our
market makers (same for all) and denote by $u_m=(u_m(x))_{x\in
\mathbf{R}}$ market maker $m$'s utility function for terminal
wealth.

%as2.1 #&#
\begin{Assumption}
\label{as1}
Each $u_m = u_m(x)$, $m=1,\ldots,M$, is a strictly concave, strictly
increasing, continuously differentiable, and bounded from above
function on the real line $\mathbf{R}$ satisfying
%
%e2.1 #&#
\begin{equation}
\label{eq7} \lim_{x\to\infty} u_m(x) = 0.
\end{equation}
\end{Assumption}

The normalizing condition (\ref{eq7}) is added only for notational
convenience. Our main results will be derived under the following
additional condition on the utility functions, which, in particular,
implies their boundedness from above.

%as2.2 #&#
\begin{Assumption}
\label{as2}
Each utility function $u_m = u_m(x)$, $m=1,\ldots,M$, is twice
continuously differentiable and its absolute risk aversion
coefficient is bounded away from zero and infinity, that is, for
some $c>0$,
\[
\frac{1}c \leq a_m(x) \set-\frac{u_m''(x)}{u_m'(x)} \leq c, \qquad
x \in\mathbf{R}.
\]
\end{Assumption}

The prices quoted by the market makers are also influenced by their
initial endowments $\alpha_0=(\alpha^m_0)_{m=1,\ldots,M} \in
\mathbf{L}^0(\mathbf{R}^M)$, where $\alpha^m_0$ is an
$\mathcal{F}$-measurable random variable describing the terminal
wealth of the $m$th market maker (if the large investor, introduced
later, will not trade at all). We assume that the initial allocation
$\alpha_0$ is \emph{Pareto optimal} in the sense of:

%de2.3 #&#
\begin{Definition}
\label{def1}
Let $\mathcal{G}$ be a $\sigma$-field contained in $\mathcal{F}$. A
vector of \mbox{$\mathcal{F}$-}\break measurable random variables
$\alpha=(\alpha^m)_{m=1,\ldots,M}$ is called \emph{a Pareto optimal
allocation given the information $\mathcal{G}$} or just a
\emph{$\mathcal{G}$-Pareto allocation} if
%
%e2.2 #&#
\begin{equation}
\label{eq8} \mathbb{E}\bigl[\bigl\llvert u_m\bigl(
\alpha^m\bigr)\bigr\rrvert \mid\mathcal{G}\bigr] < \infty, \qquad m=1,\ldots,M,
\end{equation}
and there is no other allocation $\beta\in
\mathbf{L}^0(\mathbf{R}^M)$ with the same total endowment,
%
%e2.3 #&#
\begin{equation}
\label{eq9} \sum_{m=1}^M
\beta^m=\sum_{m=1}^M
\alpha^m,
\end{equation}
leaving all market makers not worse and at least one of them better
off in the sense that
%
%e2.4 #&#
\begin{equation}
\label{eq10} \mathbb{E}\bigl[{u_m\bigl(\beta^m\bigr)}
\mid\mathcal{G}\bigr] \geq \mathbb{E}\bigl[{u_m\bigl(
\alpha^m\bigr)}\mid\mathcal{G}\bigr]\qquad\mbox{for all } m=1,\ldots,M
\end{equation}
and
%
%e2.5 #&#
\begin{equation}\label{eq11}
\quad\mathbb{P}\bigl[\mathbb{E}\bigl[{u_m\bigl(
\beta^m\bigr)}\mid\mathcal{G}\bigr] > \mathbb{E}\bigl[{u_m
\bigl(\alpha^m\bigr)}\mid\mathcal{G}\bigr]\bigr] > 0\qquad\mbox{for
some } m \in \lbrace{1,\ldots,M} \rbrace.
\end{equation}
A Pareto optimal allocation given the trivial $\sigma$-field
$\mathcal{F}_0$ is simply called a \emph{Pareto allocation}.
\end{Definition}

In other words, Pareto optimality is a stability requirement for an
allocation of wealth which ensures that there are no mutually
beneficial trades that can be struck between market makers.

Finally, we consider an economic agent or investor who is going to
trade dynamically in the financial market formed by a bank account and
$J$ stocks. We assume that the interest rate on the bank account is
given \emph{exogenously} and is not affected by the investor's trades;
for simplicity of notation, we set it to be zero. The stocks pay
terminal dividends $\psi=(\psi^j)_{j=1,\ldots,J} \in
\mathbf{L}^0(\mathbf{R}^J)$. Their prices are computed
\emph{endogenously} and depend on investor's order flow.

As the result of trading with the investor, up to and including time
$t\in[0,T]$, the total endowment of the market makers may change from
$\Sigma_0 \set\sum_{m=1}^M \alpha^m_0$ to
%
%e2.6 #&#
\begin{equation}
\label{eq12} \Sigma(\xi,\theta) \set\Sigma_0 + \xi+ \langle\theta,\psi \rangle =\Sigma_0 + \xi+ \sum_{j=1}^J
\theta^j \psi^j,
\end{equation}
where $\xi\in\mathbf{L}^0(\mathcal{F}_t,\mathbf{R})$ and $\theta
\in
\mathbf{L}^0(\mathcal{F}_t,\mathbf{R}^J)$ are, respectively, the cash
amount and the number of assets \emph{acquired} by the market makers
from the investor; they are $\mathcal{F}_t$-measurable random
variables with values in $\mathbf{R}$ and $\mathbf{R}^J$,
respectively. Our model will assume that $\Sigma(\xi,\theta)$ is
allocated among the market makers in the form of an
$\mathcal{F}_t$-Pareto allocation. For this to be possible, we have to
impose:

%as2.4 #&#
\begin{Assumption}
\label{as3}
For every $x\in\mathbf{R}$ and $q\in\mathbf{R}^J$, there is an
allocation $\beta\in\mathbf{L}^0(\mathbf{R}^M)$ with total random
endowment $\Sigma(x,q)$ defined in (\ref{eq12}) such that
%
%e2.7 #&#
\begin{equation}
\label{eq13} \mathbb{E}\bigl[u_m\bigl(\beta^m\bigr)
\bigr] > - \infty, \qquad m=1,\ldots,M.
\end{equation}
\end{Assumption}

See (\ref{eq35}) for an equivalent reformulation of this assumption
in terms of the aggregate utility function. For later use, we verify
its conditional version.

%le2.5 #&#
\begin{Lemma}
\label{lem1}
Under Assumptions~\ref{as1} and~\ref{as3}, for every
$\sigma$-field $\mathcal{G}\subset\mathcal{F}$ and random variables
$\xi\in\mathbf{L}^0(\mathcal{G},\mathbf{R})$ and $\theta\in
\mathbf{L}^0(\mathcal{G},\mathbf{R}^J)$ there is an allocation
$\beta\in\mathbf{L}^0(\mathbf{R}^M)$ with total endowment
$\Sigma(\xi,\theta)$ such that
%
%e2.8 #&#
\begin{equation}
\label{eq14} \mathbb{E}\bigl[u_m\bigl(\beta^m\bigr)
\mid\mathcal{G}\bigr] > - \infty, \qquad m=1,\ldots,M.
\end{equation}
\end{Lemma}

\begin{pf}
Clearly, it is sufficient to verify~(\ref{eq14}) on each of the
$\mathcal{G}$-measurable sets
\[
A_n \set \bigl\lbrace{\omega\in\Omega}\dvtx  \bigl\llvert \xi(\omega )
\bigr\rrvert + \bigl\llvert \theta(\omega)\bigr\rrvert \leq n \bigr\rbrace, \qquad
n\geq1,
\]
which shows that without loss of generality we can assume $\xi$ and
$\theta$ to be bounded when proving~(\ref{eq14}). Then
$(\xi,\theta)$ can be written as a convex combination of finitely
many points $(x_k,q_k) \in\mathbf{R}^{1+J}$, $k=1,\ldots,K$ with
$\mathcal{G}$-measurable weights $\lambda^k \geq0$, $\sum_{k=1}^K
\lambda^k=1$. By Assumption~\ref{as3}, for each $k=1,\ldots,K$
there is an allocation $\beta_k$ with the total endowment
$\Sigma(x_k,q_k)$ such that
\[
\mathbb{E}\bigl[u_m\bigl(\beta^m_k\bigr)
\bigr] > - \infty, \qquad m=1,\ldots,M.
\]
Thus, the allocation
\[
\beta\set\sum_{k=1}^K
\lambda^k \beta_k
\]
has the total endowment $\Sigma(\xi,\theta)$ and, by the concavity
of the utility functions, satisfies (\ref{eq13}), and hence,
also~(\ref{eq14}).
\end{pf}

%s2.2 #&#
\subsection{Simple strategies}\label{secsimplestrategies}

An investment strategy of the agent is described by a predictable
$J$-dimensional process $Q = (Q_t)_{0\leq t\leq T}$, where $Q_t =\break 
(Q_t^j)_{j=1,\ldots,J}$ is the cumulative number of the stocks
\emph{sold} by the investor through his transactions up to time
$t$. For a strategy to be self-financing we have to complement $Q$ by
a corresponding predictable process $X = (X_t)_{0\leq t\leq T}$
describing the cumulative amount of cash \emph{spent} by the
investor. Hereafter, we shall call such an $X$ a \emph{cash balance}
process.

%re2.6 #&#
\begin{Remark}
\label{rem1}
Our description of a trading strategy follows the standard practice
of mathematical finance except for the sign: positive values of $Q$
or $X$ now mean \emph{short} positions for the investor in stocks or
cash, and hence total \emph{long} positions for the market
makers. This convention makes future notation more simple and
intuitive.
\end{Remark}

To facilitate the understanding of the economic assumptions behind our
model, we consider first the case of a simple strategy $Q$ where
trading occurs only at a finite number of times, that is,
%
%e2.9 #&#
\begin{equation}
\label{eq15} Q_t = \sum_{n=1}^{N}
\theta_n 1_{(\tau_{n-1},\tau_n]}(t), \qquad0 \leq t \leq T,
\end{equation}
with stopping times $0= \tau_0 \leq\cdots\leq\tau_N = T$ and random
variables $\theta_n \in\break 
\mathbf{L}^0(\mathcal{F}_{\tau_{n-1}},\mathbf{R}^J)$, $n=1, \ldots,
N$. It is natural to expect that, for such a strategy $Q$, the cash
balance process $X$ has a similar form:
%
%e2.10 #&#
\begin{equation}
\label{eq16} X_t = \sum_{n=1}^{N}
\xi_n 1_{(\tau_{n-1},\tau_n]}(t), \qquad0 \leq t \leq T,
\end{equation}
with $\xi_n \in\mathbf{L}^0(\mathcal{F}_{\tau_{n-1}},\mathbf{R})$,
$n=1,\ldots,N$. In our model, these cash amounts will be determined
by (forward) induction along with a sequence of conditionally Pareto
optimal allocations $(\alpha_n)_{n=1,\ldots,N}$ such that each
$\alpha_n$ is an $\mathcal{F}_{\tau_{n-1}}$-Pareto allocation with the
total endowment
\[
\Sigma(\xi_n, \theta_n) = \Sigma_0 +
\xi_n + \langle\theta _n,\psi \rangle.
\]

Recall that at time $0$, before any trade with the investor has taken
place, the market makers have the initial Pareto allocation $\alpha_0$
and the total endowment $\Sigma_0$. After the first transaction of
$\theta_1$ stocks and $\xi_1$ in cash, the total random endowment
becomes $\Sigma(\xi_1,\theta_1)$. The central assumptions of our
model, which will allow us to identify the cash amount $\xi_1$
uniquely, are that, as a result of the trade:
\begin{longlist}[2.]
\item[1.] The random endowment $\Sigma(\xi_1,\theta_1)$ is redistributed
between the market makers to form a new \emph{Pareto} allocation
$\alpha_1$.
\item[2.] The market makers' expected utilities \emph{do not change}:
\[
\mathbb{E}\bigl[u_m\bigl(\alpha^m_1\bigr)
\bigr] = \mathbb{E}\bigl[u_m\bigl(\alpha^m_0
\bigr)\bigr], \qquad m=1,\ldots,M.
\]
\end{longlist}

Proceeding by induction, we arrive at the re-balance time $\tau_n$ with
the economy characterized by an $\mathcal{F}_{\tau_{n-1}}$-Pareto
allocation $\alpha_n$ of the random endowment $\Sigma(\xi_n,
\theta_n)$. We assume that after exchanging $\theta_{n+1}-\theta_n$
securities and $\xi_{n+1}-\xi_{n}$ in cash the market makers will hold
an $\mathcal{F}_{\tau_n}$-Pareto allocation $\alpha_{n+1}$ of
$\Sigma(\xi_{n+1}, \theta_{n+1})$ satisfying the key condition of the
preservation of expected utilities:
%
%e2.11 #&#
\begin{equation}
\label{eq17} \mathbb{E}\bigl[u_m\bigl(\alpha^m_{n+1}
\bigr)\mid\mathcal{F}_{\tau_n}\bigr] = \mathbb{E}\bigl[u_m\bigl(
\alpha^m_n\bigr)\mid\mathcal{F}_{\tau_n}\bigr],
\qquad m=1, \ldots, M.
\end{equation}

The fact that this inductive procedure indeed works is ensured by the
following result, established in a single-period framework in~\cite{BankKram13a}, Theorem~2.6.

%th2.7 #&#
\begin{Theorem}
\label{th1}
Under Assumptions~\ref{as1} and~\ref{as3}, every sequence of stock
positions $(\theta_n)_{n=1,\ldots,N}$ as in~(\ref{eq15}) yields a
unique sequence of cash balances $(\xi_n)_{n=1,\ldots,N}$ as
in~(\ref{eq16}) and a unique sequence of allocations
$(\alpha_n)_{n=1,\ldots,N}$ such that, for each $n=1,\ldots, N$,
$\alpha_n$ is an $\mathcal{F}_{\tau_{n-1}}$-Pareto allocation of
$\Sigma(\xi_n,\theta_n)$ preserving the market makers' expected
utilities in the sense of (\ref{eq17}).
\end{Theorem}

\begin{pf}
The proof follows from Lemma~\ref{lem1} above, Lemma~\ref{lem2} below and a
standard induction argument.
\end{pf}

%le2.8 #&#
\begin{Lemma}
\label{lem2}
Let Assumption~\ref{as1} hold and consider a $\sigma$-field
$\mathcal{G}\subset\mathcal{F}$ and random variables $\gamma\in
\mathbf{L}^0(\mathcal{G},(-\infty,0)^M)$ and $\Sigma\in
\mathbf{L}^0(\mathbf{R})$. Suppose there is an allocation $\beta\in
\mathbf{L}^0(\mathbf{R}^M)$ which has the total endowment $\Sigma$
and satisfies the integrability condition~(\ref{eq14}).

Then there are a unique $\xi\in
\mathbf{L}^0(\mathcal{G},\mathbf{R})$ and a unique
$\mathcal{G}$-Pareto allocation $\alpha$ with the total endowment
$\Sigma+ \xi$ such that
\[
\mathbb{E}\bigl[u_m\bigl(\alpha^m\bigr)\mid\mathcal{G}
\bigr] = \gamma^m, \qquad m=1, \ldots,M.
\]
\end{Lemma}
\begin{pf} The uniqueness of such $\xi$ and $\alpha$ is a
consequence of the definition of the $\mathcal{G}$-Pareto optimality
and the strict concavity and monotonicity of the utility
functions. Indeed, let $\widetilde{\xi}$ and $\widetilde{\alpha}$ be
another such pair. The allocation
\[
\beta^m \set \biggl( \widetilde{\alpha}{}^m+
\frac{\xi-\widetilde{\xi}}{M} \biggr) 1_{ \lbrace{\widetilde{\xi} < {\xi}} \rbrace} + \alpha ^m
1_{ \lbrace{\widetilde\xi\geq{\xi}} \rbrace}, \qquad m=1,\ldots,M,
\]
has the same total endowment $\Sigma+\xi$ as $\alpha$. If the
$\mathcal{G}$-measurable set $ \lbrace{\widetilde\xi< {\xi
}} \rbrace$ is
not empty, then because the utility functions $(u_m)$ are strictly
increasing, $\beta$ dominates $\alpha$ in the sense of
Definition~\ref{def1} and we get a contradiction with the
$\mathcal{G}$-Pareto optimality of $\alpha$. Hence,\vspace*{1pt} $\widetilde\xi
\geq\xi$ and then, by symmetry, $\widetilde\xi={\xi}$. In this
case, the allocation $\widetilde\beta\set
(\alpha+\widetilde{\alpha})/2$ has the same total endowment as
$\alpha$ and $\widetilde{\alpha}$. If $\widetilde\alpha\neq\alpha$
then, in view of the strict concavity of the utility functions,
$\widetilde\beta$ dominates both $\alpha$ and $\widetilde\alpha$,
contradicting their $\mathcal{G}$-Pareto optimality.

To verify the existence, we shall use a conditional version of the
argument from the proof of Theorem~2.6 in~\cite{BankKram13a}. To
facilitate references, we assume hereafter that $\llvert \gamma
\rrvert \set
\sqrt{\sum_{m=1}^M (\gamma^m)^2}$ is integrable,\vspace*{1pt} that is, $\gamma
\in\mathbf{L}^1(\mathcal{G},(-\infty,0)^M)$. This extra condition
does not restrict any generality as, if necessary, we can replace
the reference probability measure $\mathbb{P}$ with the equivalent
measure $\mathbb{Q}$ such that
\[
\frac{d\mathbb{Q}}{d\mathbb{P}} = \const\frac{1}{1 + \llvert
\gamma\rrvert }.
\]
Note that because $\gamma$ is $\mathcal{G}$-measurable this change
of measure does not affect \mbox{$\mathcal{G}$-}Pareto optimality.

For $\eta\in\mathbf{L}^0(\mathcal{G},\mathbf{R})$, denote by
$\mathcal{B}(\eta)$ the family of allocations $\beta\in
\mathbf{L}^0(\mathbf{R}^M)$ with total endowments less than or equal
to $\Sigma+ \eta$ such that
\[
\mathbb{E}\bigl[{u_m\bigl(\beta^m\bigr)}\mid\mathcal{G}
\bigr] \geq\gamma^m, \qquad m=1,\ldots,M.
\]
Since the utility functions $u_m = u_m(x)$ are increasing and
converge to $0$ as $x\to\infty$ and because there is an allocation
$\beta$ of $\Sigma$ satisfying (\ref{eq14}), the set
\[
\mathcal{H} \set \bigl\lbrace{\eta\in\mathbf{L}^0(\mathcal {G},
\mathbf{R})}\dvtx  \mathcal{B}(\eta) \neq\varnothing \bigr\rbrace
\]
is nonempty. For instance, it contains the random variable
\[
\widetilde\eta\set M \sum_{n=1}^\infty n (
\setind{A_n} - \setind{A_{n-1}}),
\]
where, for $n=0,1,\ldots,$
\[
A_n \set \bigl\lbrace{\omega\in\Omega}\dvtx  \mathbb{E}
\bigl[u_m\bigl(\beta^m + n\bigr)\mid\mathcal{G}\bigr] (
\omega)\geq\gamma^m(\omega),  m=1,\ldots,M \bigr\rbrace.
\]
Indeed, by construction, $\widetilde\eta$ is
$\mathcal{G}$-measurable and, as $A_n \uparrow\Omega$,
\[
\mathbb{E}\bigl[u_m\bigl(\beta^m +\widetilde\eta/M\bigr)
\mid\mathcal{G}\bigr] \geq \gamma^m, \qquad m=1,\ldots,M.
\]
Hence, the allocation $(\beta^m + \widetilde\eta/M)_{m=1,\ldots,M}$
belongs to $\mathcal{B}(\widetilde\eta)$.

If $\eta\in\mathcal{H}$, then the set $\mathcal{B}(\eta) \in
\mathbf{L}^0(\mathbf{R}^M)$ is convex (even with respect to
\mbox{$\mathcal{G}$-}measurable weights) by the concavity of the utility
functions. Moreover, this set is bounded in
$\mathbf{L}^0(\mathbf{R}^M)$:
\[
\lim_{z\to\infty} \sup_{\beta\in\mathcal{B}(\eta)} \mathbb{P}\bigl[
\llvert \beta\rrvert \geq z\bigr] = 0.
\]
Indeed, from the properties of utility functions in
Assumption~\ref{as1} we deduce that
\[
x^- \set\max(0,-x) \leq-\frac{u_m(x)}{u'_m(0)}, \qquad x\in \mathbf{R}.
\]
Hence, for $\beta\in\mathcal{B}(\eta)$,
\[
\mathbb{E}\bigl[\bigl(\beta^m\bigr)^-\bigr] \leq\frac{1}{u'_m(0)}
\mathbb{E}\bigl[-u_m\bigl(\beta^m\bigr)\bigr] \leq
\frac{1}{u'_m(0)} \mathbb{E}\bigl[-\gamma^m\bigr]<\infty,
\]
implying that the set $ \lbrace{((\beta^m)^-)_{m=1,\ldots,M}}\dvtx
\beta\in\mathcal{B}(\eta)  \rbrace$ is bounded in
$\mathbf{L}^1(\mathbf{R}^M)$. The boundedness of $\mathcal{B}(\eta)$
in $\mathbf{L}^0(\mathbf{R}^M)$ then follows after we recall that
\[
\sum_{m=1}^M \beta^m \leq
\Sigma+ \eta, \qquad\beta\in \mathcal{B}(\eta).
\]

Observe that if the random variables $(\eta_i)_{i=1,2}$ belong to
$\mathcal{H}$, then so does their minimum $\eta_1 \wedge\eta_2$.
It follows that there is a decreasing sequence $(\eta_n)_{n\geq1}$
in $\mathcal{H}$ such that its limit $\xi$ is less than or equal to
every element of $\mathcal{H}$. Let $\beta_n \in
\mathcal{B}(\eta_n)$, $n\geq1$. As $\beta_n \in
\mathcal{B}(\eta_1)$, the family of all possible convex combinations
of $(\beta_n)_{n\geq1}$ is bounded in
$\mathbf{L}^0(\mathbf{R}^M)$. By Lemma~\textup{A1.1} in
Delbaen and Schachermayer
\cite{DelbSch94}, we
can then choose convex combinations $\zeta_n$ of $(\beta_k)_{k\geq
n}$, $n\geq1$, converging almost surely to a random variable
$\alpha\in\mathbf{L}^0(\mathbf{R}^M)$. It is clear that
%
%e2.12 #&#
\begin{equation}
\label{eq18} \sum_{m=1}^M
\alpha^m \leq\Sigma+ \xi.
\end{equation}
Since the utility functions are bounded above and, by the convexity
of $\mathcal{B}(\eta_n)$, $\zeta_n\in\mathcal{B}(\eta_n)$, an
application of Fatou's lemma yields
%
%e2.13 #&#
\begin{equation}
\label{eq19} \quad\mathbb{E}\bigl[u_m\bigl(\alpha^m\bigr)
\mid\mathcal{G}\bigr] \geq\limsup_{n\to\infty} \mathbb{E}
\bigl[u_m\bigl(\zeta^m_n\bigr)\mid\mathcal{G}
\bigr] \geq\gamma^m, \qquad m=1,\ldots,M.
\end{equation}
It follows that $\alpha\in\mathcal{B}(\xi)$. The minimality
property of $\xi$ then immediately implies that in (\ref{eq18}) and
(\ref{eq19}) we have, in fact, equalities and that $\alpha$ is a
$\mathcal{G}$-Pareto allocation.
\end{pf}

In Section~\ref{secstrat-with-cont}, we shall prove a more
constructive version of Theorem~\ref{th1}, namely,
Theorem~\ref{th3}, where the cash balances $\xi_n$ and the Pareto
allocations $\alpha_{n}$ will be given as explicit functions of their
predecessors and of the new position $\theta_n$.

The main goal of this paper is to extend the definition of the cash
balance processes $X$ from simple to general predictable strategies
$Q$. This task has a number of similarities with the construction of a
stochastic integral with respect to a semi-martingale. In particular,
we are interested in the following questions.

%qu2.9 #&#
\begin{Question}
\label{quest1}
For simple strategies $(Q^n)_{n\geq1}$ that converge to another
simple strategy $Q$ in $\ucp$, that is, such that
%
%e2.14 #&#
\begin{equation}
\label{eq20} \bigl(Q^n-Q\bigr)^*_T \set\sup
_{0\leq t\leq T}\bigl\llvert Q^n_t -
Q_t\bigr\rrvert \to0,
\end{equation}
do the corresponding cash balance processes converge in $\ucp$ as well:
\[
\bigl(X^n - X\bigr)^*_T \to0?
\]
\end{Question}

%qu2.10 #&#
\begin{Question}
\label{quest2}
For every sequence of simple strategies $(Q^n)_{n\geq1}$ converging
in $\ucp$ to a predictable process $Q$, does the sequence $(X^n)_{n\geq
1}$ of their cash balance processes converge to a predictable
process $X$ in $\ucp$?
\end{Question}

Naturally, when we have an affirmative answer to
Question~\ref{quest2}, the process $X$ should be called the cash
balance process for the strategy $Q$. Note that a predictable process
$Q$ can be approximated by simple processes as in (\ref{eq20}) if and
only if it has LCRL (left-continuous with right limits) trajectories.

The construction of cash balance processes $X$ and processes of Pareto
allocations for general strategies $Q$ will be accomplished in
Section~\ref{secstrat-with-cont}, while the answers to Questions~\ref
{quest1} and~\ref{quest2} will be given in
Section~\ref{secappr-simple-strat}. These results rely on the
parameterization of Pareto allocations in
Section~\ref{secparam-pareto-alloc} and the properties of sample
paths of the stochastic field of aggregate utilities established in
\cite{BankKram13b} and recalled in
Section~\ref{secstoch-proc-indir}.

%s3 #&#
\section{Random fields associated with Pareto allocations}\label{secstoch-processes}

Let us collect in this section some notation and results which will
allow us to work efficiently with conditional Pareto allocations. We
first recall some terminology. For a set $A\subset\mathbf{R}^d$ a map
$\xi\dvtx   A \rightarrow\mathbf{L}^0(\mathbf{R}^n)$ is called a \emph{random
field}; $\xi$ is continuous, convex, etc., if its \emph{sample
paths} $\xi(\omega)\dvtx   A \rightarrow\mathbf{R}^n$ are continuous, convex,
etc., for all $\omega\in\Omega$. A~random field $X\dvtx   A\times[0,T]
\rightarrow\mathbf{L}^0(\mathbf{R}^n)$ is called a \emph{stochastic
field} if, for $t\in[0,T]$, $X_t \set X(\cdot,t)\dvtx   A \rightarrow
\mathbf{L}^0(\mathcal{F}_t,\mathbf{R}^n)$, that is,
the random variable $X_t$ is $\mathcal{F}_t$-measurable.

%s3.1 #&#
\subsection{Parameterization of Pareto allocations}\label{secparam-pareto-alloc}

We begin by recalling the results and notation from
\cite{BankKram13a} concerning the classical parameterization of
Pareto allocations. As usual in the theory of such allocations, a key
role is played by \emph{the aggregate utility function}
%
%e3.1 #&#
\begin{equation}
\label{eq21} r(v,x) \set\sup_{x^1 + \cdots+ x^M = x} \sum
_{m=1}^M v^m u_m
\bigl(x^m\bigr), \qquad v\in(0,\infty)^M, x\in\mathbf{R}.
\end{equation}
We shall rely on the properties of this function stated in Section~3
of \cite{BankKram13b}. In particular, $r$ is continuously
differentiable and the upper bound in~(\ref{eq21}) is attained at the
unique vector $\widehat{x} = \widehat{x}(v,x)$ in $\mathbf{R}^M$
determined by either
%
%e3.2 #&#
\begin{equation}
\label{eq22} v^m u'_m\bigl(\widehat
x^m\bigr) = \frac{\partial r}{\partial x}(v,x), \qquad m=1,\ldots,M,
\end{equation}
or, equivalently,
%
%e3.3 #&#
\begin{equation}
\label{eq23} u_m\bigl(\widehat x^m\bigr) =
\frac{\partial r}{\partial v^m}(v,x), \qquad m=1,\ldots,M.
\end{equation}

Following \cite{BankKram13a}, we denote by
%
%e3.4 #&#
\begin{equation}
\label{eq24} \mathbf{A} \set(0,\infty)^M \times\mathbf{R} \times
\mathbf{R}^J,
\end{equation}
the parameter set of Pareto allocations in our economy. An element
$a\in\mathbf{A}$ will often be represented as $a=(v,x,q)$. Here,
$v\in
(0,\infty)^M$ is a Pareto weight and $x\in\mathbf{R}$ and $q\in
\mathbf{R}^J$ stand for, respectively, a cash amount and a number of
stocks owned collectively by the market makers.

According to Lemma~3.2 in \cite{BankKram13a}, for $a=(v,x,q)\in
\mathbf{A}$, the random vector $\pi(a) \in\mathbf{L}^0(\mathbf{R}^M)$
defined by
%
%e3.5 #&#
\begin{equation}
\label{eq25} v^m u'_m\bigl(
\pi^m(a)\bigr) = \frac{\partial{r}}{\partial x}\bigl(v,\Sigma (x,q)\bigr),\qquad
m=1,\ldots,M,
\end{equation}
forms a Pareto allocation and, conversely, for $(x,q)\in
\mathbf{R}\times\mathbf{R}^J$, every Pareto allocation of the total
endowment $\Sigma(x,q)$ is given by (\ref{eq25}) for some $v\in
(0,\infty)^M$. Moreover, $\pi(v_1,x,q) = \pi(v_2,x,q)$ if and only if
$v_1=cv_2$ for some constant $c>0$ and, therefore, (\ref{eq25})
defines a one-to-one correspondence between the Pareto allocations
with total endowment $\Sigma(x,q)$ and the set
\[
\mathbf{S}^M \set \Biggl\lbrace{w \in(0,1)^M}\dvtx  \sum
_{m=1}^M w^m = 1 \Biggr\rbrace,
\]
the interior of the simplex in $\mathbf{R}^M$. Following
\cite{BankKram13a}, we denote by
\[
\pi\dvtx  \mathbf{A} \rightarrow\mathbf{L}^0\bigl(\mathbf{R}^M
\bigr),
\]
the random field of Pareto allocations given by
(\ref{eq25}). Clearly, the sample paths of this random field are
continuous. From the equivalence of~(\ref{eq22}) and~(\ref{eq23}), we
deduce that the Pareto allocation $\pi(a)$ can be equivalently defined
by
%
%e3.6 #&#
\begin{equation}
\label{eq26} u_m\bigl(\pi^m(a)\bigr) =
\frac{\partial r}{\partial v^m}\bigl(v,\Sigma(x,q)\bigr), \qquad m=1,\ldots,M.
\end{equation}

In Corollary~\ref{cor1} below, we provide the description of the
conditional Pareto allocations in our economy, which is analogous
to~(\ref{eq25}). The proof of this corollary relies on the following
general and well-known fact, which is a conditional version of
Theorem~3.1 in \cite{BankKram13a}.

%th3.1 #&#
\begin{Theorem}
\label{th2}
Consider the family of market makers with utility functions
$(u_m)_{m=1,\ldots,M}$ satisfying Assumption~\ref{as1}. Let
$\mathcal{G}\subset\mathcal{F}$ be a $\sigma$-field and $\alpha\in
\mathbf{L}^0(\mathbf{R}^M)$. Then the following statements are
equivalent:
\begin{longlist}[2.]
\item[1.] The allocation $\alpha$ is $\mathcal{G}$-Pareto optimal.
\item[2.] Integrability condition (\ref{eq8}) holds and there is a
$\mathcal{G}$-measurable random variable $\lambda$ with values in
$\mathbf{S}^M$ such that
%
%e3.7 #&#
\begin{equation}
\label{eq27} \lambda^m u_m'\bigl(
\alpha^m\bigr) = \frac{\partial r}{\partial x} (\lambda, \Sigma), \qquad m=1,\ldots,M,
\end{equation}
where $\Sigma\set\sum_{m=1}^M \alpha^m$ and the function $r =
r(v,x)$ is defined in (\ref{eq21}).
\end{longlist}
Moreover, such a random variable $\lambda$ is defined uniquely in
$\mathbf{L}^0(\mathcal{G},\mathbf{S}^M)$.
\end{Theorem}

\begin{pf} $1 \Longrightarrow 2$: It is enough to show that
%
%e3.8 #&#
\begin{equation}
\label{eq28} \frac{u'_m(\alpha^m)}{u'_1(\alpha^1)} \in \mathbf{L}^0\bigl(
\mathcal{G},(0,\infty)\bigr), \qquad m = 1,\ldots,M.
\end{equation}
Indeed, in this case, define
\[
\lambda^m \set\frac{1/u'_m(\alpha^m)}{\sum_{k=1}^M 1/u'_k(\alpha^k)}, \qquad m=1,\ldots,M,
\]
and observe that, as $u'_m$ are strictly decreasing functions,
$(\alpha^m)$ is the only allocation of $\Sigma$ such that
\[
\lambda^m u'_m\bigl(\alpha^m
\bigr) = \lambda^1 u'_1\bigl(
\alpha^1\bigr), \qquad m=1,\ldots,M.
\]
However, in view of~(\ref{eq22}), an allocation with such property
is provided by~(\ref{eq27}).

Clearly, every $\lambda\in\mathbf{L}^0(\mathcal{G},\mathbf{S}^M)$
obeying (\ref{eq27}) also satisfies the equality above and, hence,
is defined uniquely.

Suppose (\ref{eq28}) fails to hold for some index $m$, for example,
for $m=2$. Then we can find a random variable $\xi$ such that
%
%e3.9 #&#
\begin{equation}
\label{eq29} \llvert \xi\rrvert \leq1, \qquad \bigl(u'_1
\bigl(\alpha^1-1\bigr) + u'_2\bigl(
\alpha^2-1\bigr)\bigr)\llvert \xi\rrvert \in\mathbf
{L}^1(\mathbf{R}),
\end{equation}
and the set
\[
A \set \bigl\lbrace{\omega\in\Omega}\dvtx  \mathbb{E}\bigl[u'_1
\bigl(\alpha ^1\bigr)\xi\mid\mathcal{G}\bigr](\omega) < 0 <
\mathbb{E}\bigl[u'_2\bigl(\alpha^2\bigr)
\xi \mid\mathcal{G}\bigr](\omega) \bigr\rbrace
\]
has positive probability. For instance, we can take
\[
\xi\set\frac{\zeta-
\widetilde{\mathbb{E}}[\zeta\mid\mathcal{G}]}{1+u'_1(\alpha^1-1)
+ u'_2(\alpha^2-1)},
\]
where
\[
\zeta\set\frac{u'_2(\alpha^2)}{u'_1(\alpha^1) + u'_2(\alpha^2)}
\]
and $\widetilde{\mathbb{E}}$ is the expectation under the
probability measure $\widetilde{\mathbb{P}}$ with the density
\[
\frac{d\widetilde{\mathbb{P}}}{d\mathbb{P}} = \const \frac{u'_1(\alpha^1)+u'_2(\alpha^2)}{1+u'_1(\alpha^1-1) +
u'_2(\alpha^2-1)}.
\]
Indeed, in this case, (\ref{eq29}) holds easily, while, as direct
computations show
\[
A = \bigl\lbrace{\omega\in\Omega}\dvtx  \widetilde{\mathbb {E}}\bigl[\bigl(\zeta-
\widetilde{\mathbb{E}}\bigl[\zeta\mid \mathcal {G}\bigr]
\bigr)^2\mid \mathcal{G}\bigr](\omega)>0 \bigr\rbrace
\]
and $\mathbb{P}[A]>0$ because $\zeta$ is not
$\mathcal{G}$-measurable.

From the continuity of the first derivatives of the utility
functions, we deduce the existence of $0<\varepsilon<1$ such that the
set
\[
B \set \bigl\lbrace{\omega\in\Omega}\dvtx  \mathbb{E}\bigl[u'_1
\bigl(\alpha^1 - \varepsilon\xi\bigr)\xi\mid\mathcal{G}\bigr](\omega)
< 0 < \mathbb {E}\bigl[u'_2\bigl(\alpha^2
+ \varepsilon\xi\bigr)\xi\mid\mathcal{G}\bigr](\omega) \bigr\rbrace
\]
also has positive probability. Denoting $\eta\set\varepsilon\xi
1_B$ and observing that, by the concavity of utility functions,
\begin{eqnarray*}
u_1\bigl(\alpha^1\bigr) &\leq& u_1\bigl(
\alpha^1 - \eta\bigr) + u'_1\bigl(
\alpha^1 - \eta\bigr)\eta,
\\
u_2\bigl(\alpha^2\bigr) &\leq& u_2\bigl(
\alpha^2 + \eta\bigr) - u'_2\bigl(
\alpha^2 + \eta\bigr)\eta,
\end{eqnarray*}
we obtain that the allocation
\[
\beta^1 = \alpha^1-\eta, \qquad\beta^2 =
\alpha^2+\eta, \qquad\beta^m = \alpha^m,
\qquad m=3,\ldots,M,
\]
satisfies (\ref{eq9}), (\ref{eq10}) and (\ref{eq11}), thus
contradicting the $\mathcal{G}$-Pareto optimality of $\alpha$.

$2 \Longrightarrow 1$: For every allocation $\beta\in
\mathbf{L}^0(\mathbf{R}^M)$ with the same total endowment $\Sigma$
as~$\alpha$, we have
%
%e3.10 #&#
\begin{equation}
\label{eq30} \sum_{m=1}^M
\lambda^m u_m\bigl(\beta^m\bigr) \leq r(
\lambda,\Sigma) = \sum_{m=1}^M
\lambda^m u_m\bigl(\alpha^m\bigr),
\end{equation}
where the last equality is equivalent to (\ref{eq27}) in view
of~(\ref{eq22}). Granted integrability as in (\ref{eq8}), this
clearly implies the $\mathcal{G}$-Pareto optimality of $\alpha$.\vadjust{\goodbreak}
\end{pf}

From Theorem~\ref{th2} and the definition of the random field
$\pi=\pi(a)$ in~(\ref{eq25}), we obtain
the following corollary.

%co3.2 #&#
\begin{Corollary}
\label{cor1}
Let Assumptions~\ref{as1} and~\ref{as3} hold and consider a
$\sigma$-field $\mathcal{G}\subset\mathcal{F}$ and random variables
$\xi\in\mathbf{L}^0(\mathcal{G},\mathbf{R})$ and $\theta\in
\mathbf{L}^0(\mathcal{G},\mathbf{R}^J)$.

Then for every $\lambda\in\mathbf{L}^0(\mathcal{G},(0,\infty)^M)$
the random vector $\pi(\lambda, \xi, \theta)$ forms a
\mbox{$\mathcal{G}$-}Pareto allocation. Conversely, every
$\mathcal{G}$-Pareto allocation of the total endowment $\Sigma(\xi,
\theta)$ is given by $\pi(\lambda, \xi, \theta)$ for some $\lambda
\in\mathbf{L}^0(\mathcal{G},(0,\infty)^M)$.
\end{Corollary}

\begin{pf}
The only delicate point is to show that the allocation
\[
\alpha^m \set\pi^m(\lambda, \xi, \theta), \qquad m=1,\ldots,M,
\]
satisfies the integrability
condition~(\ref{eq8}). Lemma~\ref{lem1} implies the existence of
an allocation $\beta$ of $\Sigma(\xi, \theta)$
satisfying~(\ref{eq14}). The result now follows from
inequality~(\ref{eq30}) which holds true by the properties of $r =
r(v,x)$.
\end{pf}

%s3.2 #&#
\subsection{Stochastic field of aggregate utilities and its conjugate}\label{secstoch-proc-indir}

A key role in the construction of the general investment strategies
will be played by the stochastic field $F$ of aggregate utilities and
its saddle conjugate stochastic field $G$ given by
%
%e3.11 #&#
%e3.12 #&#
\begin{eqnarray}
\label{eq31} F_t(a) &\set&\mathbb{E}\bigl[r\bigl(v,\Sigma(x,q)
\bigr)\mid\mathcal{F}_t\bigr], \qquad a=(v,x,q)\in\mathbf{A},
\\
G_t(b) &\set&\sup_{v\in(0,\infty)^M}\inf
_{x\in
\mathbf{R}}\bigl[ \langle v,u \rangle+ xy - F_t(v,x,q)
\bigr],
\nonumber\\[-10pt]\label{eq32} \\[-12pt]
\eqntext{b=(u,y,q)\in \mathbf{B},}
\end{eqnarray}
where $t\in[0,T]$, the aggregate utility function $r=r(v,x)$ is given
by (\ref{eq21}), the parameter set $\mathbf{A}$ is defined
in~(\ref{eq24}), and
\[
\mathbf{B} \set(-\infty,0)^M \times(0,\infty) \times
\mathbf{R}^J.
\]
These stochastic fields are studied in \cite{BankKram13b}. For the
convenience of future references, we recall below some of their
properties.

First, we need to introduce some notation. For a nonnegative integer
$m$ and an open subset $U$ of $\mathbf{R}^d$ denote by $\mathbf{C}^m =
\mathbf{C}^m(U)$ the Fr\'echet space of $m$-times continuously
differentiable maps $f\dvtx   U \rightarrow\mathbf{R}$ with the topology
generated by the semi-norms
%
%e3.13 #&#
\begin{equation}
\label{eq33} \llVert f\rrVert _{m,C} \set\sum
_{0\leq\llvert  k\rrvert \leq m} \sup_{x\in C} \bigl\llvert
D^k f(x)\bigr\rrvert.
\end{equation}
Here, $C$ is a compact subset of $U$, $k = (k_1,\ldots,k_d)$ is a
multi-index of nonnegative integers, $\llvert  k\rrvert \set
\sum_{i=1}^d k_i$,
and
%
%e3.14 #&#
\begin{equation}
\label{eq34} D^k \set \frac{\partial^{|k|}}{\partial x_1^{k_1}\cdots\partial
x_d^{k_d}}.
\end{equation}
In particular, for $m=0$, $D^{0}$ is the identity operator and
$\llVert  f\rrVert _{0,C} \set\sup_{x\in C} |f(x)|$.

For a metric space $\mathbf{X,}$ we denote by
$\mathbf{D}([0,T],\mathbf{X})$ the space of RCLL (right-continuous
with left limits) maps of $[0,T]$ to $\mathbf{X}$.

Suppose now that Assumptions~\ref{as1} and \ref{as3} hold. Note that
in \cite{BankKram13b} instead of Assumption~\ref{as3} we used the
equivalent condition:
%
%e3.15 #&#
\begin{equation}
\label{eq35} \mathbb{E}\bigl[r\bigl(v,\Sigma(x,q)\bigr)\bigr] > -\infty,
\qquad (v,x,q) \in\mathbf{A};
\end{equation}
see Lemma~3.2 in \cite{BankKram13a} for the proof of equivalence.
Theorem~4.1 and Corollary~4.3 in \cite{BankKram13b} describe in
detail the properties of the sample paths of the stochastic fields $F$
and $G$. In particular, these sample paths belong to
$\mathbf{D}([0,T],\mathbf{C}^1)$ and for every $t\in[0,T]$, $a=
(w,x,q) \in\mathbf{S}^M \times\mathbf{R}\times\mathbf{R}^J$, and
$b=(u,1,q)$ with $u\in(-\infty,0)^M$ we have the invertibility
relations
%
%e3.16 #&#
%e3.17 #&#
%e3.18 #&#
\begin{eqnarray}
\label{eq36} w &=& {\frac{\partial G_t}{\partial u}\biggl(\frac{\partial F_t}{\partial v}
(a,t),1,q\biggr)}\Big/\Biggl(\sum_{m=1}^M \frac{\partial G_t}{\partial u^m}\biggl(\frac{\partial F_t}{\partial v}(a),1,q\biggr)\Biggr),
\\
\label{eq37} x &=& G_t\biggl(\frac{\partial F_t}{\partial v}(a),1,q\biggr),
\\
u &=& \frac{\partial F_t}{\partial v}\biggl(\frac{\partial G_t}{\partial
u}(b),G(b),q\biggr)
\nonumber\\[-8pt]\label{eq38} \\[-8pt]\nonumber
&=& \frac{\partial F_t}{\partial
v} \Biggl(
{\frac{\partial G_t}{\partial u}(b)}\Big/\Biggl(\sum_{m=1}^M
\frac{\partial G}{\partial u^m}(b)\Biggr),G(b),q \Biggr).
\end{eqnarray}
Moreover, the left-limits $F_{t-}(\cdot)$ and $G_{t-}(\cdot)$ are
conjugate to each other in a sense analogous to~(\ref{eq32}) and they
also satisfy the corresponding versions of the invertibility relations
(\ref{eq36})--(\ref{eq38}).

Theorem~4.1 in \cite{BankKram13b} also states that
\[
\frac{\partial F_t}{\partial a^i}(a) = \mathbb{E}\biggl[\frac{\partial F_T}{\partial a^i}(a)\Big|
\mathcal{F}_t\biggr], \qquad t\in[0,T], a\in\mathbf{A},
\]
which, in view of~(\ref{eq26}), implies that the derivatives of $F$
with respect to $v$ equal to the \emph{expected utilities} of the
market markers given the Pareto allocation $\pi(a)$:
%
%e3.19 #&#
\begin{equation}
\label{eq39} \frac{\partial F_t}{\partial v^m}(a) = \mathbb{E}\bigl[u_m\bigl(
\pi^m(a)\bigr)\mid\mathcal{F}_t\bigr],\qquad m=1,\ldots,M.
\end{equation}
By (\ref{eq37}), the random variable $G_t(u,1,q)$ then defines the
\emph{collective cash amount} of the market makers at time $t$ when
their current expected utilities are given by $u$ and they jointly own
$q$ stocks.

If Assumption~\ref{as2} holds as well, then by Theorem~4.2 in
\cite{BankKram13b}, the sample paths of $F$ and $G$ get an extra
degree of smoothness; they now belong to
$\mathbf{D}([0,T],\mathbf{C}^2)$.

%s4 #&#
\section{Continuous-time strategies}\label{secstrat-with-cont}

We proceed now with the main topic of the paper, which is the
construction of trading strategies with general continuous-time
dynamics. Recall that the key economic assumption of our model is that
the large investor can re-balance his portfolio \emph{without changing
the expected utilities of the market makers}.

%s4.1 #&#
\subsection{Simple strategies revisited}\label{secrevis-simple-strat}

To facilitate the transition from the discrete evolution in
Section~\ref{secsimplestrategies} to the continuous dynamics below,
we begin by revisiting the case of a simple strategy
%
%e4.1 #&#
\begin{equation}
\label{eq40} Q_t = \sum_{n=1}^{N}
\theta_n 1_{(\tau_{n-1},\tau_n]}(t), \qquad0 \leq t \leq T,
\end{equation}
with stopping times $0= \tau_0 \leq\cdots\leq\tau_N = T$ and random
variables $\theta_n \in\break 
\mathbf{L}^0(\mathcal{F}_{\tau_{n-1}},\mathbf{R}^J)$, $n=1, \ldots,
N$.

The following result is an improvement over Theorem~\ref{th1} in the
sense that the forward induction for cash balances and Pareto optimal
allocations is now made explicit through the use of the
parameterization $\pi= \pi(a)$ of Pareto allocations
from~(\ref{eq25}) and the stochastic fields $F=F_t(a) = F(a,t)$ and
$G=G_t(b) = G(b,t)$ defined in~(\ref{eq31}) and~(\ref{eq32}).

Denote by $\lambda_0\in\mathbf{S}^M$ the weight of the initial Pareto
allocation $\alpha_0$. This weight is uniquely determined by
Theorem~\ref{th2}.

%th4.1 #&#
\begin{Theorem}
\label{th3}
Let Assumptions~\ref{as1} and~\ref{as3} hold and consider a simple
strategy $Q$ given by (\ref{eq40}). Then the sequence of
conditionally Pareto optimal allocations $(\alpha_n)_{n=0,\ldots,N}$
constructed in Theorem~\ref{th1} takes the form
%
%e4.2 #&#
\begin{equation}
\label{eq41} \alpha_n = \pi(\zeta_n), \qquad n=0,\ldots,N,
\end{equation}
where $\zeta_0 \set(\lambda_0,0,0)$ and the random vectors $\zeta_n
\set(\lambda_n, \xi_n,\theta_n) \in\mathbf{L}^0(\mathbf{S}^M
\times\mathbf{R} \times\mathbf{R}^J,\mathcal{F}_{\tau_{n-1}})$,
$n=1,\ldots,N$, with $\lambda_n$ and $\xi_n$ uniquely determined by
%
%e4.3 #&#
%e4.4 #&#
\begin{eqnarray}
\lambda_n &=& {\frac{\partial G}{\partial u}\biggl(\frac{\partial F}{\partial v}(\zeta_{n-1},\tau_{n-1}),1,
\theta_n,\tau_{n-1}\biggr)}
\nonumber\\[-8pt]\label{eq42}  \\[-8pt]\nonumber
&&{} \Big/
\Biggl(\sum_{m=1}^M \frac{\partial G}{\partial u^m}\biggl(\frac{\partial F}{\partial v}(\zeta_{n-1},\tau_{n-1}),1, \theta_n,\tau_{n-1}\biggr)\Biggr),
\\
\label{eq43} \xi_n &=& G\biggl(\frac{\partial F}{\partial
v}(
\zeta_{n-1},\tau_{n-1}),1,\theta_n,
\tau_{n-1}\biggr).
\end{eqnarray}
\end{Theorem}

\begin{pf}
The recurrence relations~(\ref{eq42}) and~(\ref{eq43}) clearly
determine $\lambda_n$ and $\xi_n$, $n=1,\ldots, N$, uniquely. In
view of the identity (\ref{eq39}), for conditionally Pareto optimal
allocations $(\alpha_n)_{n=0,\ldots,N}$ defined by~(\ref{eq41}) the
indifference condition (\ref{eq17}) can be expressed as
%
%e4.5 #&#
\begin{equation}
\label{eq44} \frac{\partial F}{\partial v}(\zeta_n,\tau_{n-1}) =
\frac{\partial F}{\partial v}(\zeta_{n-1},\tau_{n-1}), \qquad n=1,\ldots,N,
\end{equation}
which, by the invertibility relations~(\ref{eq36})
and~(\ref{eq37}) and the fact that $\lambda_n$ has values in
$\mathbf{S}^M$, is, in turn, equivalent to (\ref{eq42}) and
(\ref{eq43}).
\end{pf}

In the setting of Theorem~\ref{th3}, let $A\set(W,X,Q)$, where
%
%e4.6 #&#
%e4.7 #&#
\begin{eqnarray}
\label{eq45} W_t &=& \lambda_0 \setind{[0]}(t) + \sum
_{n=1}^N \lambda_n \setind{(
\tau_{n-1},\tau_n]}(t),
\\
\label{eq46} X_t &=& \sum_{n=1}^{N}
\xi_n \setind{(\tau_{n-1},\tau_n]}(t).
\end{eqnarray}
Then $A$ is a simple predictable process with values in $\mathbf{A}$:
%
%e4.8 #&#
\begin{equation}
\label{eq47} A_t = \zeta_0 \setind{[0]}(t) + \sum
_{n=1}^{N} \zeta_n
1_{(\tau_{n-1},\tau_n]}(t), \qquad0 \leq t \leq T,
\end{equation}
with $\zeta_n$ belonging to
$\mathbf{L}^0(\mathcal{F}_{\tau_{n-1}},\mathbf{A})$ and defined in
Theorem~\ref{th3}. It was shown in the proof of this theorem that the
main condition (\ref{eq17}) of the preservation of expected utilities
is equivalent to (\ref{eq44}). Observe now that (\ref{eq44}) can
also be expressed as
%
%e4.9 #&#
\begin{equation}
\label{eq48} \frac{\partial F}{\partial v}(A_t,t) = \frac{\partial F}{\partial
v}(A_0,0)
+ \int_0^t \frac{\partial F}{\partial v}(A_s,ds),
\qquad0\leq t\leq T,
\end{equation}
where, for a \emph{simple} process $A$ as in (\ref{eq47}),
\[
\int_0^t \frac{\partial F}{\partial v}(A_s,ds)
\set \sum_{n=1}^N \biggl(
\frac{\partial F}{\partial v}(\zeta_n,\tau _n\wedge t) -
\frac{\partial F}{\partial v}(\zeta_n,\tau_{n-1}\wedge t) \biggr)
\]
denotes its nonlinear stochastic integral against the random field
$\frac{\partial F}{\partial v}$. Note that, contrary to (\ref{eq17})
and (\ref{eq44}), the condition (\ref{eq48}) also makes sense for
predictable processes $A$ which are not necessarily simple, provided
that the nonlinear stochastic integral $\int\frac{\partial
F}{\partial v}(A_s,ds)$ is well defined. This will be key for
extending our model to general predictable strategies in the next
section.

%s4.2 #&#
\subsection{Extension to general predictable strategies}\label{secgeneral-case}

For a general predictable process $A$, the construction of $\int
\frac{\partial F}{\partial v}(A_s,ds)$ requires additional conditions
on the stochastic field $\frac{\partial F}{\partial v} =
\frac{\partial F}{\partial v}(a,t)$; see, for example, Sznitman \cite{Szn81}
and Kunita \cite{Kunit90}, Section~3.2. We choose to rely on
\cite{Kunit90}, where the corresponding theory of stochastic
integration is developed for continuous semi-martingales. To simplify
notation, we shall work in a finite-dimensional Brownian setting. We
assume that, for every $a\in\mathbf{A}$, the martingale $F(a)$ of
(\ref{eq31}) admits an integral representation of the form
%
%e4.10 #&#
\begin{equation}
\label{eq49} F_t(a) = F_0(a) + \int
_0^t H_s(a) \,dB_s,
\qquad0\leq t\leq T,
\end{equation}
where $B$ is a $d$-dimensional Brownian motion and $H(a)$ is a
predictable process with values in $\mathbf{R}^{d}$. Of course, the
integral representation~(\ref{eq49}) holds automatically if the
filtration $(\mathcal{F}_t)_{0 \leq t \leq T}$ is generated by $B$. To
use the construction of the stochastic integral $\int\frac{\partial
F}{\partial v}(A_s,ds)$ from \cite{Kunit90}, we have to impose an
additional regularity condition on the integrand $H$ with respect to
the parameter~$a$.

%as4.2 #&#
\begin{Assumption}
\label{as4}
There exists a stochastic field $H = H_t(a)$ such that for every
$a\in\mathbf{A}$ the process $H(a)=(H_t(a))_{t\in[0,T]}$ is
predictable and satisfies the integral representation
(\ref{eq49}). In addition, for every $t\in[0,T]$, the random field
$H_t(\cdot)$ has sample paths in
$\mathbf{C}^1(\mathbf{A},\mathbf{R}^{d})$, and for every compact set
$C\subset\mathbf{A}$
\[
\int_0^T \llVert H_t\rrVert
^2_{1,C} \,dt < \infty,
\]
where the semi-norm $\llVert \cdot\rrVert _{m,C}$ is given
by~(\ref{eq33}).
\end{Assumption}

See Remark~\ref{rem4} below regarding the verification of this
assumption in terms of the primal inputs to our model.

Hereafter, we shall work under Assumptions~\ref{as1}, \ref{as3}
and~\ref{as4}. For convenience of future references, we formulate an
easy corollary of the properties of the sample paths of $F$ and $G$
stated in Section~\ref{secstoch-proc-indir}. For a metric space
$\mathbf{X}$ denote by $\mathbf{C}([0,T], \mathbf{X})$, the space of
continuous maps of $[0,T]$ to $\mathbf{X}$. Recall the definition of
the Fr\'echet space $\mathbf{C}^m$ from
Section~\ref{secstoch-proc-indir}.

%le4.3 #&#
\begin{Lemma}
\label{lem3}
Under Assumptions~\ref{as1}, \ref{as3} and~\ref{as4}, the
stochastic fields $F = F_t(a)$ and $G=G_t(b)$ have sample paths in
$\mathbf{C}([0,T], \mathbf{C}^1)$. If, in addition,
Assumption~\ref{as2} holds, then $F$ and $G$ have sample paths in
$\mathbf{C}([0,T], \mathbf{C}^2)$.
\end{Lemma}

\begin{pf}
As we recalled in Section~\ref{secstoch-proc-indir}, Theorem~4.1 in
\cite{BankKram13b} implies that under Assumptions~\ref{as1} and
\ref{as3} the stochastic fields $F$ and $G$ have sample paths in
the space $\mathbf{D}([0,T], \mathbf{C}^1)$ of RCLL maps and that
their left-limits satisfy conjugacy relations analogous
to~(\ref{eq32}). Moreover, under the additional
Assumption~\ref{as2}, Theorem~4.2 in \cite{BankKram13b} implies
that the sample paths of $F$ and $G$ belong to $\mathbf{D}([0,T],
\mathbf{C}^2)$. These results readily imply the assertions of the
lemma as soon as we observe that, in view of~(\ref{eq49}), for
every $a\in\mathbf{A}$, the trajectories of the martingale $F(a)$
are continuous.
\end{pf}

We also need the following elementary fact. Recall that if $\xi$ and
$\eta$ are stochastic fields on $A$ then $\eta$ is a
\emph{modification} of $\xi$ if $\xi(x) = \eta(x)$ for every $x\in A$.

%le4.4 #&#
\begin{Lemma}
\label{lem4}
Let $m$ be a nonnegative integer, $U$ be an open set in
$\mathbf{R}^n$, and $\xi\dvtx   U \rightarrow\mathbf{L}^0(\mathbf{R})$
be a
random field with sample paths in $\mathbf{C}^m = \mathbf{C}^m(U)$
such that for every compact set $C\subset U$
%
%e4.11 #&#
\begin{equation}
\label{eq50} \mathbb{E}\bigl[\llVert \xi\rrVert _{m,C}\bigr] <
\infty.
\end{equation}
Assume also that there are a Brownian motion $B$ with values in
$\mathbf{R}^d$ and a stochastic field $H = H_t(x)\dvtx   U\times[0,T]
\rightarrow\mathbf{R}^d$ such that for every $t\in[0,T]$ the random
field $H_t(\cdot)$ has sample paths in
$\mathbf{C}^m(U,\mathbf{R}^d)$ and such that for every $x\in U$ the
process $H(x)$ is predictable with
%
%e4.12 #&#
\begin{equation}
\label{eq51} M_t(x) \set\mathbb{E}\bigl[\xi(x)\mid
\mathcal{F}_t\bigr] = M_0(x) + \int_0^t
H_s(x) \,dB_s.
\end{equation}
Suppose finally that for every compact set $C\subset U$
%
%e4.13 #&#
\begin{equation}
\label{eq52} \int_0^T \llVert
H_t\rrVert ^2_{m,C} \,dt < \infty.
\end{equation}
Then $M$ has a modification with sample paths in $\mathbf{C}([0,T],
\mathbf{C}^m(U))$ and for $t\in[0,T]$, $x\in U$, and a multi-index
$k=(k_1,\ldots,k_n)$ with $|k|\leq m$,
%
%e4.14 #&#
\begin{equation}
\label{eq53} D^k M_t(x) = D^k
M_0(x) + \int_0^t D^k
H_s(x) \,dB_s,
\end{equation}
where the differential operator $D^k$ is given by~(\ref{eq34}).
\end{Lemma}

\begin{pf}
Observe first that~(\ref{eq50}) implies that $M$ has a modification
with sample paths in $\mathbf{D}([0,T], \mathbf{C}^m(U))$; see
Lemma~\textup{C.1} in \cite{BankKram13b}. We shall work with this
modification. As, for every $x\in U$, the martingale $M(x)$ is
continuous, we deduce that the sample paths of $M$ belong to
$\mathbf{C}([0,T], \mathbf{C}^m(U))$.

To verify~(\ref{eq53}), it is sufficient to consider the case $m=1$
and $k=(1,0,\ldots,0)$. Denote $e_1 \set(1,0,\ldots,0)\in
\mathbf{R}^n$. By~(\ref{eq50}),
\[
\lim_{\varepsilon\to0}\mathbb{E} \biggl[\frac{1}\varepsilon\biggl
\llvert \xi(x+\varepsilon e_1) - \xi(x) - \varepsilon
\frac{\partial
\xi}{\partial x_1}(x)\biggr\rrvert \biggr] = 0
\]
and then, by Doob's inequality,
\[
\lim_{\varepsilon\to0}\frac{1}\varepsilon \biggl(M(x+\varepsilon
e_1) - M(x) - \varepsilon\frac{\partial
M}{\partial x_1}(x) \biggr)^*_T
= 0,
\]
where $X^*_T \set\sup_{t\in[0,T]} \llvert  X_t\rrvert $.
Observe also that
by~(\ref{eq52})
\[
\lim_{\varepsilon\to0}\int_0^T \biggl
\llvert \frac{1}\varepsilon \biggl(H(x+\varepsilon e_1) -
H(x) - \varepsilon\frac{\partial H}{\partial
x_1}(x)\biggr)\biggr\rrvert ^2 \,dt =
0.
\]
The result now follows from the fact that for a sequence of
continuous local martingales $(N^n)_{n\geq1}$ its maximal elements
$(N^n)^*_T \set\sup_{t\in[0,T]} \llvert  N^n_t\rrvert $
converge to $0$ in
probability if and only if the initial values $N^n_0$ and the
quadratic variations $\langle N^n \rangle_T$ converge to $0$ in
probability.
\end{pf}

%re4.5 #&#
\begin{Remark}
\label{rem2}
If the filtration is generated by a $d$-dimensional Brownian motion
$B$, then the integral representation~(\ref{eq51}) holds
automatically. In this case, the results of \cite{Szn81}, based on
Sobolev's embeddings and It\^o's isometry, show that (\ref{eq52})
with $m=m_1$ follows from~(\ref{eq50}) with $m=m_2$ provided that
$m_1<m_2-d/2$; see our companion paper~\cite{BankKram13}.
\end{Remark}

From Lemma~\ref{lem4}, we deduce
\[
\frac{\partial F_t}{\partial v}(a) = \frac{\partial F_0}{\partial
v}(a) + \int_0^t
\frac{\partial H_s}{\partial v}(a)\,dB_s.
\]
Following Section~3.2 in \cite{Kunit90}, we say that a predictable
process $A$ with values in $\mathbf{A}$ is \emph{integrable} with
respect to the \emph{kernel} $\frac{\partial F}{\partial v}(\cdot,dt)$
or, equivalently, that the stochastic integral $\int\frac{\partial
F}{\partial v}(A_s,ds)$ is \emph{well defined} if
\[
\int_0^T \biggl\llvert \frac{\partial H_t}{\partial v}(A_t)
\biggr\rrvert ^2 \,dt < \infty.
\]
In this case, we set
\[
\int_0^t \frac{\partial F}{\partial v}(A_s,ds)
\set\int_0^t \frac
{\partial
H_s}{\partial v}(A_s)
\,dB_s,\qquad0\leq t\leq T.
\]

We are now in a position to give a formal definition of a general
trading strategy. Recall that processes $X$ and $Y$ are
\emph{indistinguishable} if $(X-Y)^*_T \set\sup_{t\in
[0,T]}\llvert  X_t-Y_t\rrvert =0$.

%de4.6 #&#
\begin{Definition}
\label{def2}
A predictable process $Q$ with values in $\mathbf{R}^J$ is called a
\emph{strategy} if there are unique (in the sense of
indistinguishability) predictable processes $W$ and $X$ with values
in $\mathbf{S}^M$ and $\mathbf{R}$, respectively, such that, for
$A\set(W,X,Q)$, the initial Pareto allocation is given by
%
%e4.15 #&#
\begin{equation}
\label{eq54} \alpha_0 = \pi(A_0),
\end{equation}
the stochastic
integral $\int\frac{\partial F}{\partial v}(A_s,ds)$ is well defined
and~(\ref{eq48}) holds.
\end{Definition}

%re4.7 #&#
\begin{Remark}
\label{rem3}
From now on, the term ``strategy'' will always be used in the sense
of Definition~\ref{def2}. Note that, at this point, it is still an
open question whether a simple predictable process $Q$ is a (valid)
strategy, as in Theorem~\ref{th3} the uniqueness of $W$ and $X$,
such that $A\set(W,X,Q)$ solves (\ref{eq48}), was proved only in
the class of \emph{simple} processes. The affirmative answer to this
question will be given in Theorem~\ref{th8} below, where in
addition to the standing Assumptions~\ref{as1}, \ref{as3}
and~\ref{as4}, we shall also require Assumptions~\ref{as2}
and~\ref{as5}.
\end{Remark}

The predictable processes $W$ and $X$ in Definition~\ref{def2} will
be called the \emph{Pareto weights} and \emph{cash balance} processes
for the strategy $Q$. We remind the reader, that the bookkeeping in
our model is done from the collective point of view of the market
makers; see Remark~\ref{rem1}. In other words, for a strategy $Q$,
the number of shares and the amount of cash \emph{owned by the large
investor} at time $t$ are given by $-Q_t$ and $-X_t$.

Accounting for~(\ref{eq39}), we call
%
%e4.16 #&#
\begin{equation}
\label{eq55} U_t \set\frac{\partial F_t}{\partial v}(A_t), \qquad0
\leq t\leq T,
\end{equation}
the process of \emph{expected utilities} for the market makers.
Observe that, as $U<0$ and $U-U_0$ is a stochastic integral with
respect to a Brownian motion, $U$ is a local martingale and a (global)
sub-martingale. The invertibility relations~(\ref{eq36}) and
(\ref{eq37}) imply the following expressions for $W$ and $X$ in terms
of $U$ and $Q$:
%
%e4.17 #&#
%e4.18 #&#
\begin{eqnarray}
\label{eq56} W_t &=& {\frac{\partial G_t}{\partial u}(U_t,1,Q_t)}
\Big/\Biggl(\sum_{m=1}^M
\frac{\partial G_t}{\partial u^m}(U_t,1,Q_t)\Biggr),
\\
\label{eq57} X_t &=& G_t(U_t,1,Q_t).
\end{eqnarray}

We also call
%
%e4.19 #&#
\begin{equation}
\label{eq58} V_t \set-G_t(U_t,1,0) =
-G_t\biggl(\frac{\partial F_t}{\partial v}(A_t),1,0\biggr), \qquad0
\leq t\leq T,
\end{equation}
the \emph{cumulative gain} process for the large trader. This term is
justified as, by~(\ref{eq57}), $V_t$ represents the cash amount the
agent will hold at $t$ if he liquidates his position in stocks. Of
course, at maturity
\[
V_T = -\bigl(X_T + \langle Q_T,\psi \rangle
\bigr).
\]

It is interesting to observe that, contrary to the standard, small
agent, model of mathematical finance, no further ``admissibility''
conditions on a strategy $Q$ are needed to exclude an arbitrage.

%le4.8 #&#
\begin{Lemma}
\label{lem5}
Let Assumptions~\ref{as1}, \ref{as3} and~\ref{as4} hold and $Q$
be a strategy such that the terminal gain of the large trader is
nonnegative: $V_T\geq0$. Then, in fact, $V_T=0$.
\end{Lemma}

\begin{pf}
Recall the notation $\lambda_0\in\mathbf{S}^M$ for the weights and
$\Sigma_0 \in\mathbf{L}^0(\mathbf{R}^M)$ for the total endowment of
the initial Pareto allocation $\alpha_0$ and $r = r(v,x)$ for the
aggregate utility function from~(\ref{eq21}). Denote by $\alpha_1$
the terminal wealth distribution between the market makers at
maturity resulting from the strategy $Q$. From the characterization
of Pareto allocations in Theorem~\ref{th2} and the sub-martingale
property of the process $U$ of expected utilities, we obtain
\begin{eqnarray*}
\mathbb{E}\bigl[r(\lambda_0, \Sigma_0)\bigr] &=&
\mathbb{E}\Biggl[\sum_{m=1}^m
\lambda^m_0 u_m\bigl(\alpha^m_0
\bigr)\Biggr] = \langle\lambda_0,U_0 \rangle\leq
\mathbb{E}\bigl[ \langle\lambda_0,U_T \rangle\bigr]
\\
&=& \mathbb{E}\Biggl[\sum_{m=1}^m
\lambda^m_0 u_m\bigl(\alpha^m_1
\bigr)\Biggr] \leq \mathbb{E}\bigl[r(\lambda_0, \Sigma_0 -
V_T)\bigr].
\end{eqnarray*}
Since $r(\lambda_0,\cdot)$ is a strictly increasing function, the
result follows.
\end{pf}

We state now a key result of the paper where we reduce the question
whether a predictable process $Q$ is a strategy to the unique
solvability of a stochastic differential equation parameterized by
$Q$.

%th4.9 #&#
\begin{Theorem}
\label{th4}
Under Assumptions~\ref{as1}, \ref{as3} and~\ref{as4}, a
predictable process $Q$ with values in $\mathbf{R}^J$ is a strategy
if and only if the stochastic differential equation
%
%e4.20 #&#
\begin{equation}
\label{eq59} U_t = U_0 + \int_0^t
K_s(U_s,Q_s) \,dB_s,
\end{equation}
has a unique strong solution $U$ with values in $(-\infty,0)^M$ on
$[0,T]$, where
\[
U^m_0 \set\mathbb{E}\bigl[u_m\bigl(
\alpha^m_0\bigr)\bigr], \qquad m=1,\ldots,M,
\]
and, for $u \in(-\infty,0)^M$, $q \in\mathbf{R}^J$ and $t \in
[0,T]$,
%
%e4.21 #&#
\begin{equation}
\label{eq60} K_t(u,q) \set\frac{\partial H_t}{\partial v}\biggl(
\frac{\partial
G_t}{\partial
u}(u,1,q),G_t(u,1,q),q\biggr).
\end{equation}
In this case, $U$ is the process of expected utilities, and the
processes of Pareto weights $W$ and cash balance $X$ are given
by~(\ref{eq56}) and~(\ref{eq57}).
\end{Theorem}

\begin{pf}
Observe that the stochastic field $F = F_t(v,x,q)$ is positive
homogeneous with respect to $v$:
\[
F_t(cv,x,q) = cF_t(v,x,q), \qquad c>0,
\]
and that the integrand $H=H_t(v,x,q)$, clearly, shares same
property. It follows that
\[
\frac{\partial H_t}{\partial v}(cv,x,q) = \frac{\partial
H_t}{\partial
v}(v,x,q), \qquad c>0,
\]
and, therefore, that the stochastic field $K$ from~(\ref{eq60}) can
also be written as
\[
K_t(u,q) = \frac{\partial H_t}{\partial
v} \Biggl(
{\frac{\partial G_t}{\partial u}(u,1,q)}
\Big/\Biggl(\sum_{m=1}^M\frac{\partial G_t}{\partial u^m}(u,1,q)\Biggr),G_t(u,1,q),q
\Biggr).
\]
After this observation, the result is an immediate consequence of
the definition of a strategy and the expressions~(\ref{eq56}) and
(\ref{eq57}) for the processes of Pareto weights $W$ and cash
balance $X$.
\end{pf}

%re4.10 #&#
\begin{Remark}
\label{rem4}
In the follow-up paper \cite{BankKram13}, we provide sufficient
conditions for a locally bounded predictable process $Q$ with values
in $\mathbf{R}^J$ to be a strategy, or equivalently,
for~(\ref{eq59}) to have a unique strong solution, in terms of the
``original'' inputs to the model: the utility functions
$(u_m)_{m=1,\ldots,M}$, the initial endowment $\Sigma_0$, and the
dividends $\psi$. In particular, these conditions also imply
Assumptions~\ref{as4} and~\ref{as5} on $H=H_t(a)$.
\end{Remark}

As an illustration, we give an example where~(\ref{eq59}) is a linear
equation, and, hence, can be solved explicitly.

%ex4.11 #&#
\begin{Example}[(Bachelier model with price impact)]
\label{ex1}
Consider an economy with a single market maker and one stock. The
market maker's utility function is exponential:
\[
u(x) = -\frac{1}{\gamma} e^{-\gamma x}, \qquad x\in\mathbf{R},
\]
where the constant $\gamma>0$ is the absolute risk-aversion
coefficient. The initial endowment of the market maker and the
payoff of the stock are given by
\begin{eqnarray*}
\Sigma_0 &=& \alpha_0 = b + \frac{\mu}{\gamma\sigma}
B_T,
\\
\psi&=& s + \mu T + \sigma B_T,
\end{eqnarray*}
where the constants $b,\mu,s \in\mathbf{R}$ and $\sigma>0$. Note
that the initial Pareto pricing measure $\mathbb{Q}=\mathbb{Q}_0$
and the stock price $S$ have the expressions
\begin{eqnarray*}
\frac{d\mathbb{Q}}{d\mathbb{P}} &\set&\const u'(\Sigma_0) =
e^{-(\mu/\sigma)B_T - (\mu^2/(2\sigma^2))T},
\\
S_t &\set& \mathbb{E}_{\mathbb{Q}_0}[\psi\mid\mathcal{F}_t]
= s + \mu t + \sigma B_t, \qquad t\in[0,T],
\end{eqnarray*}
and coincide with the martingale measure and the stock price in the
classical Bachelier model for a ``small'' investor.

Direct computations show that, for $a=(v,x,q)\in\mathbf{A}$,
\[
F_t(a) = v e^{-\gamma x} N_t(q),
\]
where the martingale $N(q)$ evolves as
%
%e4.22 #&#
\begin{equation}
\label{eq61} dN_t(q) = - \biggl(\frac{\mu}{\sigma} + \gamma\sigma
q\biggr) N_t(q) \,dB_t.
\end{equation}
For the integrand $H=H_t(a)$ in~(\ref{eq49}) and the stochastic
field $G =G_t(b)$, we obtain
\begin{eqnarray*}
\frac{\partial H_t}{\partial v}(a) &=& - \biggl(\frac{\mu}{\sigma} + \gamma\sigma q\biggr)
e^{-\gamma x} N_t(q),
\\
u &=& e^{-\gamma G_t(u,1,q)} N_t(q), \qquad u\in(-\infty,0),
\end{eqnarray*}
where the second equality follows from~(\ref{eq38}). The stochastic
field $K=K_t(u,q)$ in~(\ref{eq60}) is then given by
\[
K_t(u,q) = - \biggl(\frac{\mu}{\sigma} + \gamma\sigma q\biggr) u,
\qquad u\in (-\infty,0).
\]

From Theorem~\ref{th4}, we obtain that a predictable process $Q$ is
a strategy if and only if
\[
\int_0^T Q^2_t \,dt <
\infty,
\]
and that, in this case, the expected utility process $U$ for the
market maker evolves as
%
%e4.23 #&#
\begin{equation}
\label{eq62} dU_t = - \biggl(\frac{\mu}{\sigma} + \gamma\sigma
Q_t\biggr) U_t \,dB_t.
\end{equation}

Observe now that, by~(\ref{eq58}), the cumulative gain $V_t$ of the
large trader satisfies
\[
U_t = e^{\gamma V_t} N_t(0).
\]
From~(\ref{eq61}) and~(\ref{eq62}) and the fact that $V_0=0$, we
deduce
\begin{eqnarray*}
V_t &=& \int_0^t \biggl[
(-Q_r) (\mu \,dr + \sigma \,dB_r) - \frac{\gamma\sigma^2}{2}Q^2_r
\,dr \biggr]
\\
&=& \int_0^t \biggl[ (-Q_r)
\,dS_r - \frac{\gamma\sigma^2}{2}Q^2_r \,dr \biggr].
\end{eqnarray*}
Recall that $-Q$ denotes the number of shares owned by the large
investor and then observe that the first, linear with respect to
$Q$, term yields the wealth evolution in the classical Bachelier
model. The second, quadratic, term thus describes the feedback
effect of the large trader's actions on stock prices, with the
risk-aversion coefficient $\gamma>0$ playing the role of a
\emph{price impact coefficient}.
\end{Example}

%s4.3 #&#
\subsection{Maximal local strategies}\label{seclocal-strateg}

For a stochastic process $X$ and a stopping time $\sigma$ with values
in $[0,T]$, recall the notation $X^\sigma\set
(X_{t\wedge\sigma})_{0\leq t\leq T}$ for $X$ ``stopped'' at
$\sigma$. The following localization fact for strategies will be used
later on several occasions.

%le4.12 #&#
\begin{Lemma}
\label{lem6}
Let Assumptions~\ref{as1}, \ref{as3} and~\ref{as4} hold,
$\sigma$ be a stopping time with values in $[0,T]$, $Q$ be a
strategy and $W$, $X$, $V$ and $U$ be its processes of Pareto
weights, cash balance, cumulative gain and expected utilities. Then
$Q^\sigma$ is also a strategy and $W^\sigma$ and $X^\sigma$ are its
processes of Pareto weights and cash balance. The processes of
cumulative gain, $V(Q^\sigma)$, and of expected utilities,
$U(Q^\sigma)$, for the strategy $Q^\sigma$ coincide with $V$ and $U$
on $[0,\sigma]$, while on $(\sigma,T]$ they are given~by
\begin{eqnarray*}
U\bigl(Q^\sigma\bigr)_t &=& \frac{\partial F_t}{\partial v}(W_\sigma,X_\sigma,Q_\sigma),
\\
V\bigl(Q^\sigma\bigr)_t &=& -G_t\bigl(U
\bigl(Q^\sigma\bigr)_t,1,0\bigr).
\end{eqnarray*}
\end{Lemma}

\begin{pf}
The proof follows directly from Definition~\ref{def2} and the construction of
$U$ and $V$ in~(\ref{eq55}) and~(\ref{eq58}).
\end{pf}

Let $\tau$ be a stopping time with values in
$(0,T]\cup \lbrace{\infty} \rbrace$ and $U$ be a process
with values in
$(-\infty,0)^M$ defined on $[0,\tau)\cap[0,T]$. Recall that, for the
equation~(\ref{eq59}), $\tau$ and $U$ are called the \emph{explosion}
time and the \emph{maximal local solution} if for every stopping time
$\sigma$ with values in $[0,\tau)\cap[0,T]$ the process $U^\sigma$ is
the unique solution to (\ref{eq59}) on $[0,\sigma]$ and
%
%e4.24 #&#
\begin{equation}
\label{eq63} \limsup_{t\uparrow\tau} \bigl\llvert
\log(-U_t)\bigr\rrvert = \infty\qquad\mbox{on } \lbrace{\tau< \infty}
\rbrace.
\end{equation}
Observe that, for $m=1,\ldots,M$, the sub-martingale property of $U^m<0$
insures the existence of the limit: $\lim_{t\uparrow\tau} U^m_t$ and
prevents it from being $-\infty$. Hence, (\ref{eq63}) is equivalent
to
\[
\lim_{t\uparrow\tau} \max_{m=1,\ldots,M}U^m_t
= 0\qquad\mbox{on } \{\tau< \infty\}.
\]

For convenience of future references, we introduce a similar localized
concept for strategies.

%de4.13 #&#
\begin{Definition}
\label{def3}
A predictable process $Q$ with values in $\mathbf{R}^J$ is called a
\emph{maximal local strategy} if there are a stopping time $\tau$
with values in $(0,T]\cup \lbrace{\infty} \rbrace$ and
processes $V$, $W$
and $X$ on $[0,\tau)\cap[0,T]$ with values in $\mathbf{R}$,
$\mathbf{S}^M$ and $\mathbf{R}$, respectively, such that
%
%e4.25 #&#
\begin{equation}
\label{eq64} \lim_{t\uparrow\tau} V_t = -\infty\qquad
\mbox{on } \lbrace {\tau< \infty} \rbrace
\end{equation}
and for every stopping time $\sigma$ with values in $[0,\tau)\cap
[0,T]$ the process $Q^\sigma$ is a strategy with Pareto weights
$W^\sigma$ and cash balance $X^\sigma$ whose cumulative gain equals
$V$ on $[0,\sigma]$.
\end{Definition}

Similar to the ``global'' case we call $V$, $W$ and $X$ from
Definition~\ref{def3} the processes of cumulative gain, Pareto
weights and cash balance, respectively; the process $U$ of expected
utilities is defined on $[0,\tau)\cap[0,T]$ as in~(\ref{eq55}). In
view of~(\ref{eq64}), we call $\tau$ the \emph{explosion} time for
$V$. Note that, by Lemma~\ref{lem6}, the class of maximal local
strategies contains the class of (global) strategies.

%th4.14 #&#
\begin{Theorem}
\label{th5}
Let Assumptions~\ref{as1}, \ref{as3} and~\ref{as4} hold and
$\tau$ be a stopping time with values in
$(0,T]\cup \lbrace{\infty} \rbrace$. A predictable
process $Q$ with values in
$\mathbf{R}^J$ is a maximal local strategy and $\tau$ is the
explosion time for its cumulative gain process $V$ if and only if
the stochastic differential equation~(\ref{eq59}) admits the unique
maximal local solution $U$ with the explosion time $\tau$.

If, in addition, $Q$ is locally bounded, then $\tau$ is also the
explosion time for its cash balance process:
\[
\lim_{t\uparrow\tau} X_t = \infty\qquad\mbox{on } \lbrace {
\tau< \infty} \rbrace.
\]
\end{Theorem}

\begin{pf}
By\vspace*{1pt} Theorem~4.1 in \cite{BankKram13b}, for every $t\in[0,T]$ the
random field $G_t(\cdot)$ has sample paths in a certain space
$\widetilde{\mathbf{G}}^1$ of continuously differentiable saddle
functions on $\mathbf{B}$. Among other properties, a function $g =
g(b) = g(u,y,q)$ in $\widetilde{\mathbf{G}}^1$ is convex with
respect to $q$, strictly increasing with respect to $u$, and
%
%e4.26 #&#
\begin{equation}
\label{eq65} \lim_{n\to\infty} g(u_n,1,q) = \infty
\end{equation}
for\vspace*{1.5pt} every sequence $(u_n)_{n\geq1}$ in $(-\infty,0)^M$ converging
to a boundary point of $(-\infty,0)^M$; see the properties (G2),
(G3) and (G6) of the elements of $\widetilde{\mathbf{G}}^1$ in~\cite{BankKram13b}.

These properties readily imply that if $(g_n)_{n\geq1}$ is a
sequence in $\widetilde{\mathbf{G}}^1$ which converges to $g\in
\widetilde{\mathbf{G}}^1$ in $\mathbf{C}^1(\mathbf{B})$, then
%
%e4.27 #&#
\begin{equation}
\label{eq66} \lim_{n\to\infty} \inf_{q\in C}
g_n(u_n,1,q) = \infty
\end{equation}
for every compact set $C\subset\mathbf{R}^J$ and every sequence
$(u_n)_{n\geq1}$ in $(-\infty,0)^M$ converging to a boundary point
of $(-\infty,0)^M$. Indeed, because of the $q$-convexity and the
$u$-monotonicity, it is sufficient to consider the case when $C$ is
a singleton and the sequence $(u_n)_{n\geq1}$ is increasing. Then,
for $q\in\mathbf{R}^J$,
\[
\liminf_{n\to\infty} g_n(u_n,1,q) \geq\lim
_{k\to\infty} \liminf_{n\to
\infty}g_n(u_k,1,q)
= \lim_{k\to\infty} g(u_k,1,q) = \infty,
\]
where the last equality follows from~(\ref{eq65}).

Since, by Lemma~\ref{lem3}, the stochastic field $G=G_t(b)$ has
sample paths in $\mathbf{C}([0,T], \mathbf{C}^1(\mathbf{B}))$, the
property~(\ref{eq66}) readily yields the result as soon as we
recall the constructions of $V$ and $X$ in~(\ref{eq58}) and
(\ref{eq57}). Observe that in the argument concerning $X$ we can
assume, by localization, that $Q$ is (globally) bounded and, hence,
takes values in some compact set $C\subset\mathbf{R}^J$.
\end{pf}

To establish the existence of a maximal local strategy or,
equivalently, the existence and uniqueness of a maximal local solution
to~(\ref{eq59}) we shall also require Assumption~\ref{as2} and a
stronger version of Assumption~\ref{as4}.

%as4.15 #&#
\begin{Assumption}
\label{as5}
For every $t\in[0,T]$, the random field $H_t(\cdot)$ from
Assumption~\ref{as4} has sample paths in
$\mathbf{C}^2(\mathbf{A},\mathbf{R}^{d})$ and, for every compact set
$C\subset\mathbf{A}$,
\[
\int_0^T \llVert H_t\rrVert
^2_{2,C} \,dt < \infty.
\]
\end{Assumption}

The role of these additional assumptions is to guarantee the local
Lipschitz property with respect to $u$ for the stochastic field $K$
in~(\ref{eq60}).

%le4.16 #&#
\begin{Lemma}
\label{lem7}
Let Assumptions~\ref{as1}, \ref{as2}, \ref{as3}, \ref{as4}
and~\ref{as5} hold and $K$ be the stochastic field defined in
(\ref{eq60}). Then for every $t\in[0,T]$ the random field
$K_t(\cdot)$ has sample paths in $\mathbf{C}^1((-\infty,0)^M \times
\mathbf{R}^J, \mathbf{R}^{M\times d})$ and, for every compact set
$C\subset(-\infty,0)^M \times\mathbf{R}^J$,
\[
\int_0^T \llVert K_t\rrVert
^2_{1,C} \,dt < \infty.
\]
\end{Lemma}
\begin{pf}
This follows from Assumption~\ref{as5} and the fact that by
Lem\-ma~\ref{lem3}, the stochastic field $G=G_t(b)$ has sample paths
in $\mathbf{C}([0,T],\mathbf{C}^2(\mathbf{B}))$.
\end{pf}

%th4.17 #&#
\begin{Theorem}
\label{th6}
Let Assumptions~\ref{as1}, \ref{as2}, \ref{as3}, \ref{as4}
and~\ref{as5} hold and $Q$ be a predictable process with values
in~$\mathbf{R}^J$ such that, for every compact set $C\subset
(-\infty,0)^M$,
%
%e4.28 #&#
\begin{equation}
\label{eq67} \int_0^T \bigl\llVert
K_t(\cdot,Q_t)\bigr\rrVert ^2_{1,C}
\,dt < \infty.
\end{equation}
Then $Q$ is a maximal local strategy.
\end{Theorem}
\begin{pf}
It is well known (see, e.g., Theorem 3.4.5 in
\cite{Kunit90}) that (\ref{eq67}) implies the existence of a
unique \emph{maximal local solution} to (\ref{eq59}). The result
now follows from Theorem~\ref{th5}.
\end{pf}

%th4.18 #&#
\begin{Theorem}
\label{th7}
Under Assumptions~\ref{as1}, \ref{as2}, \ref{as3}, \ref{as4}
and~\ref{as5} every locally bounded predictable process $Q$ is a
maximal local strategy.
\end{Theorem}
\begin{pf}
This follows from Theorem~\ref{th6} if we observe that, by
Lem\-ma~\ref{lem7}, a locally bounded $Q$ satisfies~(\ref{eq67}).
\end{pf}

The preceding result allows us to finally reconcile
Definition~\ref{def2} with the construction of simple strategies in
Theorems~\ref{th1} and~\ref{th3} since it resolves the uniqueness
issue raised in Remark~\ref{rem3}.

%th4.19 #&#
\begin{Theorem}
\label{th8}
Under Assumptions~\ref{as1}, \ref{as2}, \ref{as3}, \ref{as4}
and~\ref{as5} every simple predictable process $Q$ with values in
$\mathbf{R}^J$ is a strategy and its processes of Pareto weights $W$
and cash balance $X$ are simple and given by
(\ref{eq45})--(\ref{eq46}) and \mbox{(\ref{eq42})--(\ref{eq43})}.
\end{Theorem}
\begin{pf}
The fact, that, for $W$ and $X$ given by
(\ref{eq45})--(\ref{eq46}) and (\ref{eq42})--(\ref{eq43}), the
process $A\set(W,X,Q)$ satisfies (\ref{eq54}) and (\ref{eq48})
has been already established in our discussion following
Theorem~\ref{th3}. The uniqueness follows from Theorem~\ref{th7}.
\end{pf}

%s5 #&#
\section{Approximation by simple strategies}\label{secappr-simple-strat}

In this final section, we provide a justification for the construction
of the general strategies in Definition~\ref{def2} by discussing
approximations based on simple strategies. To simplify the
presentation, we restrict ourselves to the case of locally bounded
processes.

For measurable stochastic processes, in addition to the $\ucp$
convergence defined by the metric
\[
d_{\ucp}(X,Y) \set\mathbb{E}\Bigl[\sup_{t\in[0,T]}\llvert
X_t - Y_t\rrvert \wedge 1\Bigr],
\]
we also consider the convergence in the space
$\mathbf{L}^0(d\mathbb{P}\times dt)$ with the metric
\[
d_{\mathbf{L}^0}(X,Y) \set\mathbb{E}\biggl[\int_0^T
\bigl(\llvert X_t - Y_t\rrvert \wedge1\bigr) \,dt\biggr].
\]
We call a sequence of stochastic processes $(X^n)_{n\geq1}$
\emph{uniformly locally bounded from above} if there is an increasing
sequence of stopping times $(\sigma_n)_{n\geq1}$ such that
$\mathbb{P}[\sigma_n < T]\to0$, $n\to\infty$ and $X^k_t \leq n$ on
$[0,\sigma_n]$ for $k\geq1$. The sequence $(X^n)_{n\geq1}$ is called
\emph{uniformly locally bounded} if the sequence of its absolute
values $(\llvert  X^n\rrvert )_{n\geq1}$ is uniformly locally
bounded from
above.

We begin with a general convergence result:

%th5.1 #&#
\begin{Theorem}
\label{th9}
Let Assumptions~\ref{as1}, \ref{as2}, \ref{as3}, \ref{as4}
and~\ref{as5} hold and consider a sequence of strategies
$(Q^n)_{n\geq1}$ which is uniformly locally bounded and converges
to a strategy $Q$ in $\mathbf{L}^0(d\mathbb{P}\times dt)$.

Then the processes $(U^n,V^n)_{n\geq1}$, of expected utilities and
cumulative gains, converge to $(U,V)$ in $\ucp$, the processes
$(W^n,X^n)_{n\geq1}$, of Pareto weights and cash balance, converge
to $(W,X)$ in $\mathbf{L}^0(d\mathbb{P}\times dt)$, and the sequence
$(X^n)_{n\geq1}$ is uniformly locally bounded. If, in addition,
the sequence $(Q^n)_{n\geq1}$ converges to $Q$ in~$\ucp$, then the
sequence $(W^n,X^n)_{n\geq1}$ also converges to $(W,X)$ in $\ucp$.
\end{Theorem}
\begin{pf} By standard localization arguments, we can assume the
existence of constants $a>0$ and $b>0$ such that
\[
\max\Bigl(\bigl\llvert \ln(-U)\bigr\rrvert,\llvert Q\rrvert,\sup
_{n\geq1}\bigl\llvert Q^n\bigr\rrvert \Bigr) \leq a,
\]
and, in view of Lemma~\ref{lem7}, such that
%
%e5.1 #&#
\begin{equation}
\label{eq68} \int_0^T \bigl\llVert
K_s(\cdot)\bigr\rrVert ^2_{1,C(a)}\,ds \leq b,
\end{equation}
where
\[
C(a) \set \bigl\lbrace{(u,q)\in(-\infty,0)^M\times\mathbf
{R}^J}\dvtx  \max\bigl(\bigl\llvert \ln(-u)\bigr\rrvert, \llvert q
\rrvert \bigr) \leq2a \bigr\rbrace.
\]

Define the stopping times
\[
\sigma_n \set\inf \bigl\lbrace{t\in[0,T]}\dvtx  \bigl\llvert \ln
\bigl(-U^n_t\bigr)\bigr\rrvert \geq2a \bigr\rbrace, \qquad
n\geq1,
\]
where we follow the convention that $\inf
\varnothing\set\infty$. Observe that the $\ucp$ convergence of
$(U^n)_{n\geq1}$ to $U$ holds if
%
%e5.2 #&#
\begin{equation}
\label{eq69} \bigl(U-U^n\bigr)^*_{T\wedge\sigma_n} \to0, \qquad n
\to\infty.\vadjust{\goodbreak}
\end{equation}

To prove (\ref{eq69}), note first that for every two stopping times
$0 \leq\tau_* \leq\tau^* \leq\sigma_n$ we have using Doob's
inequality
\begin{eqnarray*}
\mathbb{E}&&\Bigl[\sup_{\tau_* \leq t \leq\tau^*} \bigl\llvert U_t-U^n_t
\bigr\rrvert ^2\Bigr]
\\
&&\qquad \leq\mathbb{E} \biggl[2 \bigl\llvert U_{\tau_*}-U^n_{\tau_*}
\bigr\rrvert ^2+2 \sup_{\tau_* \leq t
\leq\tau^*} \biggl\llvert \int
_{\tau
_*}^t\bigl(K_s(U_s,Q_s)-K_s
\bigl(U^n_s,Q^n_s\bigr)
\bigr)\,dB_s\biggr\rrvert ^2\biggr]
\\
&&\qquad \leq2 \mathbb{E} \bigl\llvert U_{\tau_*}-U^n_{\tau_*}
\bigr\rrvert ^2 +8 \mathbb{E}\biggl[\int_{\tau_*}^{\tau^*}
\bigl\llvert K_s(U_s,Q_s) -
K_s\bigl(U^n_s,Q^n_s
\bigr)\bigr\rrvert ^2\,ds\biggr]
\\
&&\qquad \leq2 \mathbb{E} \bigl\llvert U_{\tau_*}-U^n_{\tau_*}
\bigr\rrvert ^2
\\
&&\quad\qquad{} +8\mathbb{E}\biggl[\int_{\tau_*}^{\tau^*}
\bigl\llVert K_s(\cdot)\bigr\rrVert ^2_{1,C(a)}
\bigl(\bigl\llvert U_s-U^n_s\bigr\rrvert
^2 + \bigl\llvert Q_s - Q^n_s
\bigr\rrvert ^2\bigr) \,ds\biggr]
\\
&&\qquad \leq2 \mathbb{E} \bigl\llvert U_{\tau_*}-U^n_{\tau_*}
\bigr\rrvert ^2+8\mathbb{E}\biggl[\int_{\tau_*}^{\tau^*}
\bigl\llVert K_s(\cdot)\bigr\rrVert ^2_{1,C(a)}
\,ds \sup_{\tau_* \leq
t \leq\tau^*} \bigl\llvert U_t-U^n_t
\bigr\rrvert ^2\biggr]
\\
&&\quad\qquad{}+ 8 \mathbb{E}\biggl[\int_{\tau_*}^{\tau^*} \bigl
\llVert K_s(\cdot)\bigr\rrVert ^2_{1,C(a)} \bigl
\llvert Q_s - Q^n_s\bigr\rrvert
^2 \,ds\biggr].
\end{eqnarray*}
Rearranging terms, we thus obtain
%
%e5.3 #&#
%e5.4 #&#
\begin{eqnarray}\label{eq70}
&& \mathbb{E}\biggl[\biggl(1-8\int_{\tau_*}^{\tau^*}
\bigl\llVert K_s(\cdot)\bigr\rrVert ^2_{1,C(a)}
\,ds\biggr)\sup_{\tau_* \leq
t \leq\tau^*} \bigl\llvert U_t-U^n_t
\bigr\rrvert ^2\biggr]
\nonumber\\[-8pt]\\[-8pt]\nonumber
&&\qquad \leq2 \mathbb{E} \bigl\llvert U_{\tau_*}-U^n_{\tau_*}
\bigr\rrvert ^2+8 \mathbb{E}\biggl[\int_{\tau_*}^{\tau^*}
\bigl\llVert K_s(\cdot)\bigr\rrVert ^2_{1,C(a)}
\bigl\llvert Q_s - Q^n_s\bigr\rrvert
^2 \,ds\biggr].
\end{eqnarray}
Now choose $\tau_0 \set0$ and, for $i=1,2,\ldots,$ let
\[
\tau_i \set\inf \biggl\lbrace{t \geq\tau_{i-1}}\dvtx  8\int
_{\tau
_{i-1}}^t \bigl\llVert K_s(\cdot)\bigr
\rrVert ^2_{1,C(a)}\,ds \geq\frac{1}2 \biggr\rbrace
\wedge T.
\]
Note that because of (\ref{eq68}) we have $\tau_i = T$ for $i \geq
i_0$, where $i_0$ is the smallest integer greater than $16b$. Hence,
to establish (\ref{eq69}), it suffices to prove
\[
\mathbb{E}\Bigl[\sup_{\tau_{i-1} \wedge\sigma^n \leq s \leq\tau_i
\wedge\sigma^n} \bigl\llvert U_s-U^n_s
\bigr\rrvert ^2\Bigr] \to0, \qquad n\to\infty \mbox{ for } i=1,
\ldots,i_0.
\]
For $i=1$, this follows from estimate (\ref{eq70}) with $\tau_* \set
\tau_0 =0$ and $\tau^* \set\tau_1 \wedge\sigma^n$ because
$U_0=U^n_0$ and because of our assumption on the sequence $(Q^n)_{n
\geq1}$. For $i=2,3,\ldots$ this convergence holds by induction,
since with $\tau_* \set\tau_{i-1} \wedge\sigma^n$ and $\tau^*
\set
\tau_i \wedge\sigma^n$ the first term on the right-hand side of
(\ref{eq70}) vanishes for $n \to\infty$ because of the validity of
our claim for $i-1$ and the second term disappears again by
assumption on $(Q^n)_{n \geq1}$. This completes the proof of the $\ucp$
convergence of $(U^n)_{n\geq1}$ to $U$.

The rest of the assertions follows from the
representations~(\ref{eq56}), (\ref{eq57}) and~(\ref{eq58}) for
Pareto weights, cash balances and cumulative gains in terms of the
stochastic field $G = G_t(b)$ and the fact that, by
Lemma~\ref{lem3}, $G$ has sample paths in
$\mathbf{C}(\mathbf{C}^1(\mathbf{B}), [0,T])$.
\end{pf}

%th5.2 #&#
\begin{Theorem}
\label{th10}
Under Assumptions~\ref{as1}, \ref{as2}, \ref{as3}, \ref{as4}
and~\ref{as5}, a predictable locally bounded process $Q$ with
values in $\mathbf{R}^J$ is a strategy if and only if there is a
sequence $(Q^n)_{n\geq1}$ of simple strategies,\vspace*{2pt} which is uniformly
locally bounded, converges to $Q$ in $\mathbf{L}^0(d\mathbb{P}\times
dt)$, and for which the sequence of associated cash balances
$(X^n)_{n\geq1}$ is uniformly locally bounded from above.
\end{Theorem}

For the proof, we need a lemma.

%le5.3 #&#
\begin{Lemma}
\label{lem8}
Under Assumptions~\ref{as1}, \ref{as2}, \ref{as3}
and~\ref{as4}, for every strategy $Q$ and every $t\in[0,T]$
%
%e5.5 #&#
%e5.6 #&#
%e5.7 #&#
\begin{eqnarray}
\label{eq71}  && \sum_{m=1}^M \biggl(
\frac{1}c \log\bigl(\bigl(-U^m_t\bigr)\vee1\bigr)
+ c \log\bigl(\bigl(-U^m_t\bigr) \wedge 1\bigr) \biggr)\nonumber
\\
&&\qquad \leq G_t(-\idvec,1,Q_t) - X_t
\\
&&\qquad\leq\sum_{m=1}^M \biggl(
\frac{1}c \log\bigl(\bigl(-U^m_t\bigr)\wedge1
\bigr) + c \log\bigl(\bigl(-U^m_t\bigr)\vee1\bigr)
\biggr),\nonumber
\end{eqnarray}
where $c>0$ is taken from Assumption~\ref{as2}, $\idvec\set
(1,\ldots,1)\in\mathbf{R}^M$, and $X$ and $U$ are the processes of
cash balance and expected utilities for $Q$.
\end{Lemma}
\begin{pf}
Theorem~4.2 in \cite{BankKram13b} implies that under
Assumptions~\ref{as1}, \ref{as2} and~\ref{as3}, for every $t\in
[0,T]$ the random field $G_t(\cdot)$ has sample paths in a certain
space $\widetilde{\mathbf{G}}^2(c)$ of twice-differentiable saddle
functions on $\mathbf{B}$. The property (G7) of the elements of
$\widetilde{\mathbf{G}}^2(c)$ states that
\[
\frac{1}c \leq-u^m \frac{\partial G_t}{\partial u^m}(u,1,q) \leq c, \qquad
m=1,\ldots, M.
\]
This yields the result if we account for the
representation~(\ref{eq57}) for $X$.
\end{pf}

\begin{pf*}{Proof of Theorem~\ref{th10}}
The ``only if'' part follows from Theorem~\ref{th9} and the fact
that every locally bounded predictable process $Q$ can be
approximated in $\mathbf{L}^0(d\mathbb{P}\times dt)$ by a sequence
of simple predictable processes $(Q^n)_{n\geq1}$ which is uniformly
locally bounded. Hereafter, we shall focus on sufficiency.

By Theorem~\ref{th7}, $Q$ is a maximal local strategy. Denote by
$U$ and $X$ its processes of expected utilities and cash balance and
by $\tau$ the explosion time of $X$; see Theorem~\ref{th5}. We
have to show that $\tau= \infty$.

For $a>0$ and $b>a$, define the stopping times
\begin{eqnarray*}
\tau(a) &\set&\inf \Bigl\lbrace{t\in[0,T]}\dvtx  \max_{m=1,\ldots,M}
U^m_t > -a \Bigr\rbrace,
\\
\tau_n(a) &\set&\inf \Bigl\lbrace{t\in[0,T]}\dvtx  \sup
_{k\geq n} \max_{m=1,\ldots,M} U^{k,m}_t
> -a \Bigr\rbrace, \qquad n\geq1,
\\
\sigma(b) &\set&\inf \Bigl\lbrace{t\in[0,T]}\dvtx  \min_{m=1,\ldots,M}
U^m_t < -b \Bigr\rbrace,
\\
\sigma_n(b) &\set&\inf \Bigl\lbrace{t\in[0,T]}\dvtx  \inf
_{k\geq n} \min_{m=1,\ldots,M} U^{k,m}_t
< -b \Bigr\rbrace, \qquad n\geq1,
\end{eqnarray*}
where $U^n$ is the process of expected utilities for $Q^n$ and where
we let $\inf\varnothing\set\infty$. Note that, by
Theorem~\ref{th5}, $\tau(a)\to\tau$, $a\to0$, and hence, $\tau=
\infty$ if and only if
%
%e5.8 #&#
\begin{equation}
\label{eq72} \lim_{a\to0}\mathbb{P}\bigl[\tau(a) \leq T
\bigr] = 0.
\end{equation}

From Theorem~\ref{th5} and Lemma~\ref{lem6}, we deduce that
$Q^{\tau(a)\wedge T}$ is a strategy whose expected utility process
coincides with $U$ on $[0,\tau(a)\wedge T]$. Hence, by
Theorem~\ref{th9},
%
%e5.9 #&#
\begin{equation}
\label{eq73} \bigl(U^n - U\bigr)^*_{\tau(a)\wedge T}\to0, \qquad n
\to\infty.
\end{equation}
Hereafter, we shall assume that $a$ is rational and that, for every
such $a$, the convergence above takes place almost surely. This can
always be arranged by passing to a subsequence.

Since
\[
\bigl\{\tau(a) < \tau_n(2a)\bigr\} \subset\bigcap
_{k\geq n} \bigl\{\bigl(U^k-U\bigr)^*_{\tau(a)\wedge T}
\geq a\bigr\},
\]
we obtain
%
%e5.10 #&#
\begin{equation}
\label{eq74} \lim_{n\to\infty} \mathbb{P}\bigl[\tau(a) <
\tau_n(2a)\bigr] = 0.
\end{equation}
Similarly, as
\[
\bigl\{\sigma_n(2b)\wedge\tau(a) < \sigma(b)\wedge\tau(a)\bigr\}
\subset\bigcup_{k\geq n} \bigl\{\bigl(U^k-U
\bigr)^*_{\tau(a)\wedge T} \geq b\bigr\},
\]
and since the convergence in~(\ref{eq73}) takes place almost
surely, we deduce
\[
\lim_{n\to\infty} \mathbb{P}\bigl[\sigma_n(2b)\wedge
\tau(a) < \sigma (b)\wedge \tau(a)\bigr] = 0.
\]
The latter convergence implies that
%
%e5.11 #&#
\begin{equation}
\label{eq75} \limsup_{n\to\infty} \mathbb{P}\bigl[
\sigma_n(2b)<\tau(a)\bigr] \leq \mathbb{P}\bigl[\sigma(b) < \tau(a)
\bigr] \leq\mathbb{P}\bigl[\sigma(b) < \tau\bigr].
\end{equation}
From (\ref{eq74}) and (\ref{eq75}), we deduce
\[
\mathbb{P}\bigl[\tau(a)\leq T\bigr] \leq\mathbb{P}\bigl[\sigma(b) < \tau\bigr] +
\limsup_{n\to\infty}\mathbb{P}\bigl[\tau_n(2a)\leq
\sigma_n(2b)\wedge T\bigr].
\]
Therefore, (\ref{eq72}) holds if
%
%e5.12 #&#
\begin{equation}
\label{eq76} \lim_{b\to\infty} \mathbb{P}\bigl[\sigma(b) < \tau
\bigr] = 0,
\end{equation}
and, for every $b>0$,
%
%e5.13 #&#
\begin{equation}
\label{eq77} \lim_{a\to0}\limsup_{n\to\infty}
\mathbb{P}\bigl[\tau_n(a)\leq\sigma_n(b) \wedge T\bigr] =
0.
\end{equation}

The verification of (\ref{eq76}) is straightforward due to the
sub-martingale property of $U$. The uniform local boundedness
conditions on $(Q^n)_{n\geq1}$ and $(X^n)_{n\geq1}$ (from above)
and the fact that $G$ has trajectories in
$\mathbf{C}(\mathbf{C}(\mathbf{B}),[0,T])$ imply that the process
\[
Y_t \set\inf_{n\geq1} \bigl(G\bigl(-
\idvec,1,Q^n_t,t\bigr) - X^n_t
\bigr), \qquad0\leq t\leq T,
\]
is locally bounded from below. The convergence (\ref{eq77}) follows
now from the second inequality in (\ref{eq71}) of
Lemma~\ref{lem8}.
\end{pf*}

We conclude this section with affirmative answers to our
Questions~\ref{quest1} and~\ref{quest2} from
Section~\ref{secsimplestrategies}. Recall that the acronym LCRL
means left-continuous with right limits.

%th5.4 #&#
\begin{Theorem}
\label{th11}
Under Assumptions~\ref{as1}, \ref{as2}, \ref{as3}, \ref{as4}
and~\ref{as5}, a predictable process $Q$ with values in
$\mathbf{R}^J$ and LCRL trajectories is a strategy if and only if
there is a predictable process $X$ with values in $\mathbf{R}$ and a
sequence of simple strategies $(Q^n)_{n\geq1}$ converging to $Q$ in
$\ucp$ such that the sequence of its cash balances $(X^n)_{n\geq1}$
converges to $X$ in $\ucp$. In this case, $X$ is the cash balance
process for $Q$.
\end{Theorem}
\begin{pf}
This follows from Theorems~\ref{th9} and~\ref{th10} and the fact that
every predictable process with LCRL trajectories is a limit in $\ucp$
of a sequence of simple processes which then necessarily is also
uniformly locally bounded.
\end{pf}

%\begin{appendix}
%\section{}
%\end{appendix}

%\section*{Acknowledgments}

%\begin{supplement}[id=suppA]
%\slink[doi]{10.1214/00-AAPXXXXSUPP} %[doi,text={...}] - jei reikia
%suskaldyti doi
%\sfilename{aapXXXX\_supp.pdf}
%\end{supplement}

% imsref loaded by linak, 2014-09-05 13:05:56
% imsref loaded by linak, 2014-09-05 13:15:12
% imsref loaded by linak, 2014-09-05 13:20:17
% imsref loaded by linak, 2014-09-05 13:21:15
% imsref loaded by linak, 2014-09-05 13:32:53

\printaddresses
\end{document}